\documentclass[9pt]{article}
\usepackage{spconf,amsmath,epsfig}
\usepackage{amssymb}
\usepackage{amsthm}
\usepackage{graphicx}

\newtheorem{theorem}{Theorem}[section]
\newtheorem{lemma}{Lemma}[section]
\newtheorem{definition}{Definition}[section]
\newtheorem{claim}{Claim}

\newcommand{\bi}{\begin{itemize}}
\newcommand{\ei}{\end{itemize}}
\newcommand{\be}{\begin{enumerate}}
\newcommand{\ee}{\end{enumerate}}

\newcommand{\beq}{\begin{eqnarray*}}
\newcommand{\eeq}{\end{eqnarray*}}
\newcommand{\beqn}{\begin{eqnarray}}
\newcommand{\eeqn}{\end{eqnarray}}

\title{Geographic Routing with Limited Information in Sensor Networks}

\name{Sundar Subramanian and Sanjay Shakkottai\thanks{Department of
    Electrical and Computer Engineering, The University of Texas at
    Austin. This research was supported by NSF Grants ACI-0305644,
    CNS-0325788, and CNS-0347400. Email:
    \{ssubrama,shakkott\}@ece.utexas.edu.}}
\address{}

\begin{document}
\ninept
\maketitle
\begin{abstract}
Geographic routing with greedy relaying strategies have been
widely studied as a routing scheme in sensor networks. These
schemes assume that the nodes have perfect information about the
location of the destination. When the distance between the source
and destination is normalized to unity, the asymptotic routing
delays in these schemes are $\Theta(\frac{1}{M(n)}),$ where $M(n)$
is the maximum distance traveled in a single hop (transmission
range of a radio).

We consider three scenarios: {\it (i)} where nodes have location
errors (imprecise GPS), {\it (ii)} where only coarse geographic
information about the destination is available, such as the
quadrant or half-plane in which the destination is located, and
{\it (iii)} where only a small fraction of the nodes have routing
information. In this paper, we show that even with such imprecise
or limited destination-location information, the routing delays
are $\Theta(\frac{1}{M(n)})$.  We further show that routing delays
of this magnitude can be obtained even if only a small fraction of
the nodes have any location information, and other nodes simply
forward the packet to a randomly chosen neighbor, and we validate
our analysis with simulation.

Finally, we consider the throughput-capacity of networks with
progressive routing strategies that take packets closer to the
destination in every step, but not necessarily along a straight-line.
Such a routing strategy could potentially lead to spatial ``hot
spots'' in the network where many data flows intersect at a spatial
region (a node or group of nodes), due to ``sub-optimal'' routes with
increased path-lengths. In this paper, we show that the effect of hot
spots due to progressive routing does not reduce the network
throughput-capacity in an order sense. In other words, the
throughput-capacity with progressive routing is order-wise the same as
the maximum achievable throughput-capacity.

\end{abstract}
\section{Introduction}
\label{sec:int}
The availability of cheap wireless technology and the emergence of
micro-sensors based on MEMS technology will enable the ubiquitous
deployment of sensor networks
\cite{sohgaoailpot00,aksusaca02,eshegoku99}.
Applications for sensor networks include robust communication,
intrusion detection and commercial applications involving macro-scale
measurements and control.  Such networks are characterized by the
absence of any large-scale established infrastructure, and nodes
cooperate by relaying packets to ensure that the packets reach their
respective destinations.

A popular routing algorithm for sensor network that has been widely
studied is geographic routing
\cite{jaipursen01,kaku00,ingoes00,grovet03}. The main idea is to
forward a packet to a node that is closer to the final destination
than the current packet position (a greedy forwarding strategy). When
greedy forwarding fails (due to dead-ends or routing loops), alternate
routing methods such as perimeter routing, or route discovery based
methods (using flooding) have been proposed \cite{kaku00,jaipursen01}.

In this paper, we study the problem of geographic routing with
limited or erroneous destination-location information. For
instance, suppose that nodes only know the quadrant or the
half-plane on which the final destination is.
A node could then {\em randomly} forward the packet to an
arbitrary node that is in that direction. As another example,
suppose that nodes have the correct destination coordinates.
However, the GPS at nodes are erroneous (and possibly biased), as
a result of which packets are routed in the wrong direction.

\begin{figure}[t!]
\begin{center}
\includegraphics*[scale= .4]{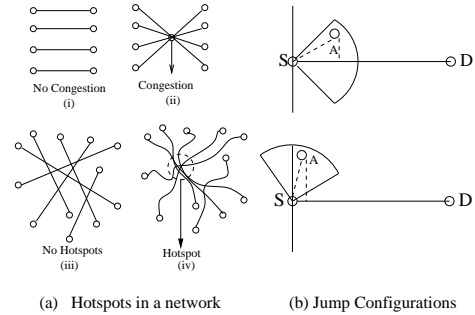}
\caption{Random Routing\label{fig:routconf}}
\end{center}
\end{figure}

\subsection{Main Contributions}

We consider a large-scale network where nodes are deployed over a
unit region. Each node's maximum transmission range is scaled as
$M(n) = K\sqrt{(\log{n}/n)},$ for some $K > 1.$ For $K$ large
enough, and $n$ large enough, results in
\cite{gupkum00,shasrishr03} ensure that straight line routing
(greedy geographic routing) is possible without recourse to face
routing (the ``loop-around'' strategy employed when straight-line
routing fails due to dead ends).

We first consider the case where nodes have precise destination
coordinates. However, we assume that the GPS at nodes are
imprecise. We model this by assuming that each routing step has an
angular error\footnote{Note that by expressing the position of a
node in polar coordinates, the radial component of the error will
not affect geographic routing; however the angular component could
point in the wrong direction.  Thus, we model GPS errors by
randomness in the angular component.}  that is random.
In other words, nodes attempt to perform greedy straight-line
routing. However, due to the angular error, the packet is
forwarded to a random node that is in some sector within angles
$\phi_1$ and $\phi_2$ (illustrated in Figure~\ref{fig:bias1}).

We then consider the case where nodes have limited destination
information. In particular, we consider the case where each node
has only a coarse estimate -- such as quadrant or half-plane
information. In other words, each node has a coordinate system (a
local notion of `North') that need not be common to all nodes. All
that each node knows is that the local quadrant in which the
destination lies (or the half-plane in which the destination
lies). In each of these cases, the routing strategy that is
adopted is to simply forward the packet to a randomly selected
node in the appropriate quadrant (or the half-plane).

We also consider the case where only a small fraction of the nodes
have any routing information at all. Most nodes simply forward the
packet to a randomly selected neighbor. A small fraction of the
nodes have quadrant information (as discussed earlier). This could
be distributed by some gossip mechanism
\cite{limama99,kermasgan01}, where nodes forward routing
information, but also forget this information after some time. We
consider a simple model where a node has routing information with
some fixed probability $p \in (0,1),$ in which case, it routes to
the appropriate quadrant and other-wise randomly routes the packet
to an arbitrary neighbor.

Finally, we consider the throughput-capacity in networks for the
special case of {\em progressive routing strategies} where the packets
are transported closer to their destinations in each step, but not
necessarily along a straight-line. Such a routing strategy could
potentially lead to spatial ``hot spots'' in the network where many
data flows intersect at a spatial region (a node or group of nodes),
due to ``sub-optimal'' routes with increased path-lengths. For
example, consider Figure~\ref{fig:routconf}(a)(i), where $n$
source-destination pairs have non-intersecting paths (each of length
'1'), thus resulting in a (normalized) throughput of '1' for each
pair. On the other hand, let us now allow the path length between each
source-destination pair to be no more than length '3' (due to imperfect
routing). Then, in the worst-case, all the paths will share a
bottle-neck node, thus decreasing the throughput per
source-destination pair to be $1/n$ (illustrated in
Figure~\ref{fig:routconf}(a)(ii)).  With $n$ randomly placed
source-destination pairs and straight-line routing (see
Figure~\ref{fig:routconf}(a)(iii)), it has been shown in
\cite{gupkum00} that the throughput-capacity per source-destination
pair scales as $1/\sqrt{n \log(n)}.$ Analogous to
Figure~\ref{fig:routconf}(a)(ii), imperfect routing strategies, even
when the path length is increased by {\em only a constant factor}
(non-straight-line routing) could lead to hot-spots, thus decreasing
throughput-capacity in an order-wise sense. In this paper, we show
that the effect of hot spots due to progressive routing does not
reduce the network throughput-capacity in an order sense. In other
words, the throughput-capacity with progressive routing is order-wise
the same as the maximum achievable throughput-capacity.

The main contributions in this paper are the following:
\begin{itemize}

\item[(i)] We show that the time to reach the destination
 with erroneous angular information or limited information
 (quadrant information) is within a constant factor of
 straight-line greedy routing. We derive upper and lower bounds on the
 routing delay which are asymptotically tight (in $n$).

\item[(ii)] We show that even in the case where only a fixed fraction
  of the nodes have routing information, the routing delay is within a
  constant factor of straight-line routing.Thus, this implies that for
  any fixed $p \in (0,1),$ we can achieve a delay within a constant
  factor of the optimal strategy. The trade-off is that the constant
  factor scales as $\frac{1}{p}.$

\item[(iii)] In the delay analysis, we adopt a continuum model of a sensor
network where packets are routed along points on the plane, and
each hop has a step-size that is bounded by $M(n)$. In
Section~\ref{sec:sim}, we validate the analytical results using
simulations where the discretization effects due to node locations
are accounted for.

\item[(iv)] For networks with progressive routing strategies, we show
  that although hot spots might occur, they are not severe enough to
  reduce the throughput-capacity in an order-wise sense.

\end{itemize}

We comment that for the strategies considered, suppose that we had a
deterministic progress toward the destination, then it is easy to see
that the routing delay\footnote{However, as discussed earlier, it is
  not clear even in this case if the throughput-capacity is unchanged
  in an order sense. We prove in Section~\ref{sec:cap} that the
  throughput-capacity does not decrease in an order sense.} will be
order-wise equivalent to straight-line routing. For example, in
Figure~\ref{fig:routconf}(b)(i), a packet from source 'S' to
destination 'D' is routed such that the packet's location at each
subsequent hop lies in a sector oriented toward the final destination
in a manner such that there is a deterministic lower-bound on the
progress toward the destination. This leads to an appropriate
deterministic upper-bound on the routing delay.

However, if a deterministic positive step does not occur, (as in
Figure~\ref{fig:routconf}(b)(ii)), then it is possible that the delay
is significantly larger. It is reasonable to expect that if the
expected distance is positive (as in (ii)), we should expect the delay
to be order-wise equivalent to straight-line routing, with a
proportionality constant equal to the inverse of the mean distance
traveled in every jump. Indeed, this would be true if the progress
toward the destination in subsequent hops were independent and
identically distributed (i.i.d.), or such that some form of the law of
large numbers were satisfied.  However in our case, the progress (the
difference between $|\overrightarrow{SD}|$ and $|\overrightarrow{AD}|$
in Figure~\ref{fig:routconf}(b)(ii)) at subsequent hops are neither
independent, nor identically distributed. In fact, the mean
progress gets smaller as we proceed towards the destination and
the sequence is correlated. We show that
even under these circumstances, we can upper and lower bound the
projections of subsequent steps by a sequence of i.i.d random
variables, and use these i.i.d. bounds to derive asymptotically tight
bounds on the routing delay.

\subsection{Related Work}

There has been considerable interest in greedy geographic routing and
the associated recovery mechanisms to route around dead-ends
\cite{kaku00,jaipursen01,krsiur99,kuwazo03}, as well as its
applications \cite{rakashesgoyiyu03}.

The idea that approximate information may be sufficiently far away from
the destination has been explored in the context of mobile ad hoc
networks. In \cite{pegech00}, the authors propose the Fish-eye state
routing, where nodes exchange link state information with a frequency
that depends on the distance from the destination.  The idea that
nodes far away from the destination requires less precise information
has been exploited in \cite{bachsywo98}, where the authors propose
lazy update mechanisms for routing tables.  In \cite{grovet03}, the
authors exploit such an effect in the context of mobile nodes to
propose Last Encounter routing, where mobile nodes remember their last
encounter time and location with other nodes. They show that with
sufficient mobility, such schemes result in a performance that is
within a constant factor of the best-case routing.
In the context of geographic routing, \cite{ampali99} have proposed a
routing protocol where a set of embedded (circular) geographic routing
zones are defined about the destination. In each zone, a packet
travels along a greedy path toward the center of the next-level zone
(a tighter circle about the destination). When it enters the next
level zone, a course correction occurs, and the packet is routed in a
greedy manner toward the center of the next-level zone. Thus, as the
packet gets closer to the destination, more detailed information is
available, leading to a sequence of course corrections. Using
simulations, the authors have shown that such a scheme is a
bandwidth-efficient routing protocol for large-scale networks.

In \cite{mepoak04}, the authors formulate the local topology knowledge
needed for optimal energy efficient geographic routing using an
integer linear program, and propose Partial Topology Forwarding
Routing. Related work also includes geographic routing with
localization errors (where a node does not know its own position
precisely). In \cite{hhbsa03}, the authors show using simulations that
localization errors of less that 0.4 times the transmission radius
does not impact the performance of greedy forwarding in geographical
routing. In \cite{sehego04}, the authors study the effect of
localization errors on face routing.  They first derive failure modes
with localization errors (such as routing loops, cross links, and
excessive edge removals). Next, using simulations, the authors in
\cite{sehego04} argue that even a 10\% localization errors can
significantly impact the performance of face routing (perimeter
routing). However, when the sensor network size is large, it has been
shown in \cite{gupkum00,shasrishr03} that with high probability,
greedy routing will succeed (i.e., recovery mechanisms such as face
routing will be required with small probability). In this paper we
study such large-scale sensor networks, and analyze the performance of
randomized-geographic routing algorithms with limited information. We
show that the delay with such schemes is asymptotically (order-wise)
equivalent to straight-line (greedy) routing. Using simulations, we
finally validate our analysis.

In Section~\ref{sec:sysmod}, we describe the system model. In
Sections~\ref{sec:nobias},\ref{sec:quad} and \ref{sec:frac}, we
derive the delay asymptotics for routing with imprecise and
limited information. In Section~\ref{sec:sim}, we present
simulation results. Finally, in Section~\ref{sec:cap}, we derive
the achievable throughput for progressive routing schemes and show
that it is order-wise equivalent to the upper bound on throughput
capacity.

\begin{figure}
    \begin{center}
        \includegraphics[scale =0.5]{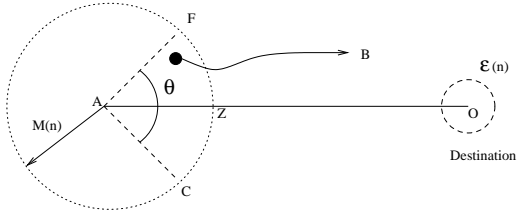}
        \caption{Routing as a hopping process\label{fig:jumps1}}
    \end{center}
\end{figure}

\section{System Description}
\label{sec:sysmod}
We consider a unit region over which sensor nodes are deployed.  All
nodes are assumed to have the same (maximum) transmission range and
can transmit to any node within its transmission radius. The
transmission regions are assumed to be circular. For a fixed $K > 1,$
We suppose that the common transmission range for all the sensors is
\beqn
M(n)&=&K \sqrt{(\log{n}/n)} \label{eqn:power_scale}
\eeqn
In this paper, we study routing behavior with limited information
in the large $n$ regime (i.e., $n \to \infty$). From results in
\cite{gupkum98,gupkum00}, such a scaling of the radius (equivalently,
the peak transmission power) leads to a sensor network with $n$
randomly placed nodes being (asymptotically) connected. Further, from
results in \cite{gupkum00,shasrishr03}, for $K$ large enough (but
finite), this scaling ensures that straight line routing (greedy
geographic routing) is possible without recourse to face routing
(the ``loop-around'' strategy employed when straight-line routing
fails due to dead ends).

For each point `A', we define its neighborhood set as the
collection of points \beqn A_{M(n)}& =& \{X \in {\cal R}^2: |
\overrightarrow{XA} | < M(n)\}, \label{eqn:neidef} \eeqn where
$|\overrightarrow{XA}|$ is the Euclidean distance between `X' and
`A'.

In this paper, we ignore the discretization effects due to node
position (see also \cite{haj00b} for a similar model). In other
words, suppose that a packet at location `A' needs to be
transmitted using geographic routing to the destination at
location `O' as in Figure~\ref{fig:jumps1}. Then, we assume that
at the next hop, the packet is routed to the point `Z' in
Figure~\ref{fig:jumps1}. For instance, suppose that the network is
a grid network with $n$ nodes over the unit square (i.e.,
$\frac{1}{\sqrt{n}}$ distance between nodes).  Then, in practice,
straight-line routing would lead to the packet at `A' being routed
to the node closest to the point `Z'. In this paper we ignore this
discretization error, as this asymptotically vanishes (the error
is at most $\frac{0.5}{\sqrt{n}},$ whereas the transmission radius
is $K \sqrt{(\log{n}/n)},$ which is order-wise larger). As another
example, suppose that we employ a random routing scheme, where the
packet at `A' needs to be routed to a randomly chosen node in the
sector `FAC' (see Figure~\ref{fig:jumps1}). Then, we assume that
the packet is forwarded to a point `B' whose location is uniformly
distributed over this sector. To summarize, we adopt a continuum
model of a sensor network where we route along points on the
plane, and each hop has a step-size that is bounded by $M(n).$ In
Section~\ref{sec:sim}, we validate the analytical results we
derive in this paper using simulations where the discretization
effects are accounted for.

We employ a two-tier routing model in this paper. We consider an
$\epsilon(n)$ ball about the destination (see Figure
~\ref{fig:jumps1}). When a packet is within this $\epsilon(n)$
ball (which is arbitrarily close to the destination, as $n$
increases), we assume that nodes have sufficient routing
information to employ straight-line routing. However, for nodes
outside this $\epsilon(n)$ ball, we consider various routing
strategies with limited information.

Our objective in this paper is to quantify the amount of routing
resources required over the network (for instance, the metric could be
the average number of bits per node in the network to route from a
randomly chosen origin to the destination).  Physically, the ball
around the destination corresponds to $\epsilon(n)$
destination-location advertisement, which is order-wise negligible in
the sense that the number of nodes in this ball is vanishingly small
when compared to the total number of nodes in the network. We also
require that the destination ball is larger than each hop step size to
overcome edge effects. Thus, we choose a suitable ball size of
$\epsilon(n)= n^{-1/4}$ in this paper (we use the parameter $1/4$ for
notational convenience; our proofs work for any radius that is
order-wise larger than each hop step-size).

With this setup, let us define $Y^{(n)}(i)$ to be the Euclidean
distance traveled towards the
destination in the $i^{th}$ step (and when the transmission range is
$M(n)$), by following the relay strategy $\pi$. We define the
routing delay $\tau(n)$ for this strategy as follows
\begin{equation}
\label{eqn:hittim} \tau(n) = \sup \big\{ j :
\sum_{i=1}^{j}Y^{(n)}(i) \le d - \epsilon(n)\big\},
\end{equation}
where $d$ is the Euclidean distance between the source
and the destination. Thus, $\tau(n)$ represents the hitting time
corresponding to a path entering the $\epsilon(n)$ ball, when the
transmission radius is $M(n).$ We say that a routing strategy $\pi$
has an order-wise straight-line routing delay if the random variable
$\tau_{\pi}(n) = \Theta(\frac{1}{M(n)})$, as this is (order-wise)
\footnote{We denote $g(n) = \Theta(f(n)),$ if there exists positive
constants $c_1$ and $c_2$ such that for all $n$ large enough, $0 < c_1
\leq g(n)/f(n) \leq c_2.$} the number of steps required for
deterministic jumps of size $M(n)$ to reach the destination a unit
distance away along a straight-line path.

\section{Analysis of Routing with Sector Information}
\label{sec:nobias}

In this section, we consider the situation where all nodes know
the destination location perfectly, but have imprecise GPS
information about their positions. This error in position
contributes to an angular error in the direction of the
destination. Hence, when the node wishes to transmit, the choice
of neighbor is not along the correct direction to the destination,
but in a sector within angles $[\phi1,\phi2]$ corresponding to the
error in angular information. The misaligned sector AFC is a
sector contained between the angles $[\phi_1, \phi_2]$, such that
for a randomly chosen point $(L,\alpha)$ from the sector,
$E(L\cos{\alpha}) >0$.

Consider Figure \ref{fig:bias1}. Let the packet be currently at
the point `A' at the $i^{th}$ step and wish to travel to the
destination `O'. An error in location is mathematically equivalent
to stating that the next hop location is randomly chosen (with an
uniform distribution) as any point in the sector AFC. The neighbor
subset from which we choose our relay node is the set
$A_{(M(n),\phi1,\phi2)}$, where
\begin{equation}
A_{(M(n),\phi1,\phi2)} = \Big\{ X \in A_{M(n)} : \phi1 <
(\angle{\overrightarrow{XA}} - \angle{\overrightarrow{OA}}) <
\phi2 \Big\}.
\end{equation}
We have assumed that the radial distance of the hop is also random,
and not deterministically equivalent to $M(n).$ The randomness in the
radial distance (per hop) models a variable power selection at the
node. The analysis in this paper can be directly extended to the case
when the radial distance is deterministic (or any other given
distribution).  $Y^{(n)}(i)$ is the Euclidean distance traveled
towards the destination in the $i^{th}$ step. By definition,
$Y^{(n)}(i) = |\overrightarrow{OA}| - |\overrightarrow{OB}|.$
We denote the polar co-ordinates of this jump as the pair
$(\hat{L}^{(n)}(i),\alpha^{(n)}(i))$,
where $\hat{L}^{(n)}(i) =
|\overrightarrow{AB}|$ and $\alpha^{(n)}(i) = \angle{OAB}$. Now,
let us consider the delay $\tau(n)$ for this routing scheme. The
packet's source is A and the destination O with
$|\overrightarrow{OA}| = 1$ for notational simplicity.
\\
\noindent \begin{definition} \label{defn:aas} We define a random
sequence \{$a(n)$,$n = 1,2,\dots $\} to be asymptotically almost
surely (a.a.s) bounded by another random sequence \{$b(n)$,$n =
1,2,\dots $\} if $\exists N_0>0$ such that for all $n > N_0$,
$a(n) \leq b(n)$ $a.s$.
\end{definition} In the rest of the paper, we denote sequences
\{$a(n)$\} and
\{$b(n)$\} satisfying Definition~\ref{defn:aas} by
\begin{displaymath}
a(n) \le b(n) \quad (a.a.s).
\end{displaymath}

We shall now show in Theorem~\ref{thm:sectheta} that the delay for
this scheme is of the order of straight-line routing. To prove
Theorem~\ref{thm:sectheta}, we will need to prove the following Lemma.

\noindent\begin{lemma}[A Limit theorem for Triangular Arrays]
\label{lem:tria}For any fixed $K > 1$, let $M(n)=K\sqrt{(\log{n}/n)}$.
Consider a triangular array of bounded i.i.d. (independent and
identically distributed) random variables
$X^{(n)}_{i}, 1\le i \le n.$ Then,
\begin{displaymath}
\lim_{n\rightarrow\infty} M(n) \sum_{i=1}^{\frac{1}{M(n)}}
X^{(n)}_{i} \longrightarrow EX^{1}_{1} \quad (a.a.s)
\end{displaymath}
\end{lemma}
 \noindent\begin{proof} We have
 \begin{eqnarray}\label{eqn:cherno}
 P\Big( |M(n) \sum_{i=1}^{\frac{1}{M(n)}}X_{i}^{(n)}-EX| >
 \epsilon\Big) < \nonumber \\
 P\Big(M(n)\sum_{i=1}^{\frac{1}{M(n)}} X_{i}^{(n)}-EX > \epsilon
 \Big) \nonumber \\
  + P\Big( M(n) \sum_{i=1}^{\frac{1}{M(n)}} X_{i}^{(n)}-EX <
 -\epsilon \Big).
 \end{eqnarray}
 By the Chernoff Bound, we have
 \begin{displaymath}
 P\Big(M(n)  \sum_{i=1}^{\frac{1}{M(n)}} X_{i}^{(n)}-EX > \epsilon
 \Big) < \exp^{-\frac{1}{M(n)}I(\epsilon)}.
 \end{displaymath}
 where $I(\epsilon)$ is the large-deviations rate function \cite{demzei98} for the
 bounded random variable. Applying the bound to $L.H.S$ of equation
 (\ref{eqn:cherno}),
 \begin{displaymath} P\Big( |M(n)
 \sum_{i=1}^{\frac{1}{M(n)}} X_{i}^{(n)}-EX| > \epsilon \Big) <
 2\exp^{-\frac{1}{M(n)}I(\epsilon)}
 \end{displaymath} Also,
 \begin{displaymath}
 2\sum_{i = 1}^{n} \exp^{-\frac{1}{M(i)}I(\epsilon)} < \infty
 ,\quad M(i) = K\sqrt{\frac{\log{i}}{i}}.
 \end{displaymath}
 Thus, by Borel-Cantelli's Lemma,
 \begin{displaymath}
 \lim_{n\rightarrow\infty} M(n) \sum_{i=1}^{\frac{1}{M(n)}}
 X^{(n)}_{i} \longrightarrow EX \quad (a.a.s).
 \end{displaymath}
 \end{proof}
It can be shown that the sequence of random variables
$\{Y^{(n)}_i\}$ are not i.i.d., but are history dependent. Thus,
we first upper and lower bound these random variables by sequence
of i.i.d. random variables, and a sequence of error terms.
\begin{lemma}\label{lem:randpointbound}
Let $(S,\alpha)$ be the polar co-ordinates of any
point B within a circle of radius $m$,
 with center A. Let O be any point on
the plane such that $|\overrightarrow{OA}| > (m+\epsilon).$ Let $
\epsilon > 0$ Then,
\begin{displaymath}
S\cos{\alpha} - \frac{S^2}{\epsilon} \le |\overrightarrow{OA}| -
|\overrightarrow{OB}| \le S\cos{\alpha}
\end{displaymath}
\end{lemma}
\begin{proof} The proof is presented in the Appendix.\end{proof}
Using the bound in Lemma~\ref{lem:randpointbound}, we now derive
the main result.
\begin{theorem}
\label{thm:sectheta} Let $(L,\alpha)$ be the polar coordinates of
a uniformly chosen point from a sector within angles
$[\phi1,\phi2]$ and unit radius. Let $\beta = E(L\cos(\alpha))$
Then, $\forall$ positive $c_1,c_2: c_1<\frac{1}{\beta}<c_2$,
\begin{displaymath}
\frac{c_1}{K}\sqrt{\frac{n}{\log(n)}} \le \tau(n) \le
\frac{c_2}{K}\sqrt{\frac{n}{\log(n)}} \quad(a.a.s).\Box
\end{displaymath}
\end{theorem}
\begin{proof} Recall that $Y^{(n)}(i)$ is the distance traveled
towards the destination in the $i^{th}$ step. For a packet located
at `A' at time-step $i$ (see Figure \ref{fig:bias1}), and the next
hop position being `B', our routing model implies that $Y^{(n)}(i)
= |\overrightarrow{OA}| - |\overrightarrow{OB}|$.

From Lemma~\ref{lem:randpointbound}, we have, for all $j <
\tau(n)$, the following equations.
\begin{equation}\label{eqn:lowbou}
\sum_{i=1}^{j}\Big\{\hat{L}^{(n)}(i)\cos{\alpha^{(n)}(i)} -
\frac{{\hat{L}^{(n)}(i)}^2}{\epsilon(n)}\Big\} < \sum_{i=1}^{j}
Y^{(n)}(i),
\end{equation}
\begin{equation}\label{eqn:upbou}
\sum_{i=1}^{j} Y^{(n)}(i) <
\sum_{i=1}^{j}\hat{L}^{(n)}(i)\cos{\alpha^{(n)}(i)}.
\end{equation}
Defining $L^{(n)}(i)= \frac{\hat{L}^{(n)}(i)}{M(n)} $ and substituting in
equations (6,7), we have
\beqn
M(n)\sum_{i=1}^{j}L^{(n)}(i)\cos{\alpha^{(n)}(i)}-&& \label{eqn:lobousca}\\
{\big(M(n)\big)}^2\sum_{i=1}^{j}\frac{{L^{(n)}(i)}^2}{\epsilon(n)}& <&
\sum_{i=1}^{j} Y^{(n)}(i) \nonumber
\eeqn
\beqn
\sum_{i=1}^{j} Y^{(n)}(i) &<& M(n)
\sum_{i=1}^{j}L^{(n)}(i)\cos{\alpha^{(n)}(i)}. \label{eqn:upbousca}
\eeqn
We observe that $\{L^{(n)}(i)\cos{\alpha^{(n)}(i)}\}_{i=1}^n$ are a
sequence of i.i.d. random variables, with the expected value
$E\{L^{(n)}(i)\cos{\alpha^{(n)}(i)}\} = \beta,$ such that $0<\beta<1$.

\emph{Upper Bound:} To prove the bounds for the hitting time, let
us suppose that our claim
$
\tau(n) \le \frac{c_2}{M(n)} \quad \forall
c_2> \frac{1}{\beta}\quad (a.a.s)
$
is not true. Then, there exists a subsequence $n_k, k = 1,2,\dots$
such that $\tau(n_k) > \frac{c_2}{M(n_k)}$. Note that, from
 (\ref{eqn:hittim}), this implies that
\begin{equation}\label{eqn:contraimplicat}
\sum_{i=1}^{\frac{c_2}{M(n_k)}} Y^{(n_k)}(i) < 1, \quad k =
1,2,\dots
\end{equation}

However, from (\ref{eqn:lobousca}), (which holds for all $j<
\tau(n)$), we have
\begin{eqnarray}\label{eqn:contra}
M(n_k)\sum_{i=1}^{\frac{c_2}{M(n_k)}}L^{(n_k)}(i)\cos{\alpha^{(n_k)}(i)} -
\nonumber \\
{\big(M(n_k)\big)}^2\sum_{i=1}^{\frac{c_2}{ M(n_k)}}
\frac{{L^{(n_k)}(i)}^2}{\epsilon(n_k)} <
\sum_{i=1}^{\frac{c_2}{M(n_k)}} Y^{(n_k)}(i)
\end{eqnarray}
By substituting $X_{i}^{(n)} = L^{(n)}(i)\cos{\alpha^{(n)}(i)}$ in
Lemma~\ref{lem:tria} and noting that almost sure
convergence along a sequence implies an almost sure
convergence along every subsequence, it follows from
Lemma~\ref{lem:tria} that
\begin{equation}\label{eqn:contraimp1}
M(n_k)\sum_{i=1}^{\frac{c_2}{M(n_k)}}
L^{(n_k)}(i)\cos{\alpha^{(n_k)}(i)} \rightarrow c_2\beta ; \quad
c_2\beta >1,
\end{equation} as $E(L^{(n_k)}(i)\cos{\alpha^{(n_k)}(i)}) = \beta$.

Moreover, since $L^{(n_k)} \le 1 $ and $\frac{M(n)}{\epsilon(n)}
\rightarrow 0$, we have
\begin{equation}\label{eqn:contraimp2}
{\big(M(n_k)\big)}^2\sum_{i=1}^{\frac{c_2}{ M(n_k)}}
\frac{{L^{(n_k)}(i)}^2}{\epsilon(n_k)} <
{\big(M(n_k)\big)}^2\sum_{i=1}^{\frac{c_2}{ M(n_k)}}
\frac{1}{\epsilon(n_k)},
\end{equation} and
\begin{equation}\label{eqn:contraimp3}
{\big(M(n_k)\big)}^2\sum_{i=1}^{\frac{c_2}{ M(n_k)}}
\frac{1}{\epsilon(n_k)} \rightarrow 0.
\end{equation} From equations
(\ref{eqn:contra},\ref{eqn:contraimp1},\ref{eqn:contraimp2} and
\ref{eqn:contraimp3}) we have
$\lim_{k\rightarrow\infty}
\sum_{i=1}^{\frac{c_2}{M(n_k)}}Y^{(n_k)}(i)$ $ > 1,$
which contradicts  (\ref{eqn:contraimplicat}).
Thus we have shown that
$\tau(n) \le \frac{c_2}{M(n)}  \quad \forall
c_2>\frac{1}{\beta}\quad (a.a.s).$

\begin{figure}
    \begin{center}
        \includegraphics[scale =0.375]{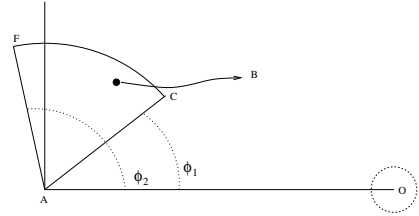}
        \caption{A sector with bias.\label{fig:bias1}}
    \end{center}
\end{figure}

\emph{Lower Bound:} To prove the lower bound, we need this
additional construction. For each $n$, let us augment the sequence
of random variables ${Y^{(n)}(i), 1 \le i \le n}$, as follows. Once
a packet has entered the $\epsilon(n)$ ball about the destination,
we start a new packet from the source to the destination. Thus we
define a sequence of random variables $Y^{(n)}(i)$ for all $i$.
These random variables generate the sequence of
${\hat{L}^{(n)}(i)\cos{\alpha^{(n)}(i)}, \forall i}$.

Let us assume that the lower bound
$\tau(n) \ge \frac{c_1}{M(n)} \quad \forall c_1 < \frac{1}{\beta}$
(a.a.s). is not true. Observe that $\tau(n) \ge
\frac{1}{M(n)}$. Then, there exists a subsequence $n_k, K =
1,2,\dots$ such that for some $r \in (1,c_1)$,
\begin{equation}\label{eqn:tau_nk_conv}
M(n_k)\tau(n_k) \rightarrow r.
\end{equation}
Let $W(n) = \frac{r}{M(n)}$. We observe that
\begin{equation*}
\Big|\sum_{i=1}^{\tau(n_k)}\hat{L}^{(n_k)}(i)\cos{\alpha^{(n_k)}(i)}
-
\sum_{i=1}^{W(n_k)}\hat{L}^{(n_k)}(i)\cos{\alpha^{(n_k)}(i)}\Big|
\end{equation*}
\begin{equation}
 \le \Big|W(n_k)-\tau(n_k)\Big|M(n_k).
\end{equation}
Thus, we have
\begin{eqnarray}\label{eqn:LowerboundIneq}
1 = \sum_{i=1}^{\tau(n_k)}Y^{(n_k)}(i) \le
\sum_{i=1}^{\tau(n_k)}\hat{L}^{(n_k)}(i)\cos{\alpha^{(n_k)}(i)}\le
\nonumber \\
\sum_{i=1}^{W(n_k)}\hat{L}^{(n_k)}(i)\cos{\alpha^{(n_k)}(i)} +
\Big|W(n_k)-\tau(n_k)\Big|M(n_k).
\end{eqnarray}
Now applying Lemma~\ref{lem:tria} and equation
(\ref{eqn:tau_nk_conv}) to  (\ref{eqn:LowerboundIneq}), we
get
\begin{eqnarray}
\sum_{i=1}^{W(n_k)}\hat{L}^{(n_k)}(i)\cos{\alpha^{(n_k)}(i)}
\rightarrow r
<1,\\
\Big|W(n_k)-\tau(n_k)\Big|M(n_k) \rightarrow 0.
\end{eqnarray} This contradicts our assumption that
$\sum_{i=1}^{\tau(n_k)}Y^{(n_k)}(i)=1$. Thus, by contradiction, we
have shown that
$\tau(n) \ge \frac{c_1}{M(n)} \quad \forall c_1 < \frac{1}{\beta}
(a.a.s).$
\end{proof}
Thus, this result implies that for large enough $n,$ the delay with
random angular error leads is equal to $\frac{1}{\beta M(n)}$ which is
clearly the same order as that with straight-line routing, with the
scaling constant inversely proportional to the expected value of the
projection of each step on the line joining the source and destination.

\section{ Routing with Quadrant Information}
\label{sec:quad}

In the previous section, we had shown that even with GPS error,
the routing delays were within a constant factor of greedy
straight-line routing. In this section, we assume that there is
some mechanism that provides coarse geographic information about
the destination, such as the quadrant or half-plane in which the
destination is located.  Under such a scenario, we derive bounds
on the routing delays. We show that even in an adversarial mode of
choosing the local quadrants, the routing delay is within a
constant factor of straight-line routing.
\begin{figure}
\begin{center}
\includegraphics*[width=0.6\columnwidth]{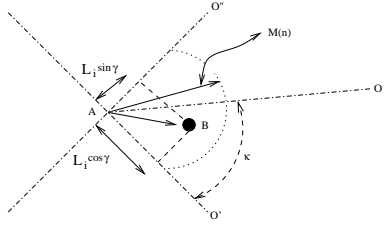}
\caption{Adversarial Quadrants -  where O' and O'' correspond to
the possible ``worst-case'' directions of the destination.
\label{fig:quad1dia}}
\end{center}
\end{figure}
Consider the following routing strategy $\Psi_2$ (see
Figure~\ref{fig:quad1dia}). The node `A' contains a packet at the
$i^{th}$ step that needs to be routed to the destination `O'. The
strategy adopted is to randomly forward the packet to a randomly
chosen point `B' from the correct quadrant.

Further, all nodes need not have the same coordinate system.  For
instance, suppose that `A' only knows that the final destination
is locally to the `North-West' (with respect to its own coordinate
system). Let us denote the offset between the node's local
coordinate system and the true direction of the destination by a
random variable $\kappa.$  We will consider two cases: {\em (i)}
the offset $\kappa$ is assumed to be uniformly distributed within
the quadrant; and {\em (ii)} an adversarial scenario where
$\kappa$ is chosen to be the worst-case at each hop, i.e., along
the one of the local coordinate axes that minimizes the distance
traveled toward the destination (see Figure~\ref{fig:quad1dia}).

Let the polar representation of `B' be
$L_{2}^{(n)}(i),\alpha^{(n)}_2(i)$. The neighbor subset from which the
relay node is chosen is given by
\begin{displaymath}
A_{\kappa,M(n)} = \{\mbox{Node }X \in A_{M(n)}: \kappa -
\frac{\pi}{2} < \angle(\overrightarrow{XA}) < \kappa \}.
\end{displaymath}

We first consider the case where the angle $\kappa= \angle{OAC}$ is
assumed to be uniformly distributed in $[0,\frac{\pi}{2}]$.
This is equivalent to picking a node 'B' from a semicircular AFC
($\phi_1 = \frac{-\pi}{2},\phi_2 = \frac{\pi}{2}$ in
Figure~\ref{fig:bias1}) with a probability distribution
\begin{equation}\label{eqn:quaddistn}
\mathbf{f_{L_2,\alpha_2}} = \frac{1}{A} \frac{\frac{\pi}{2} -
|\alpha_2|}{\frac{\pi}{2}} \quad z < M , |\alpha_2|<\frac{\pi}{2}.
\end{equation}

As before , let the source be a unit distance away from the
destination. Let us define the hitting time for the path to hit the
$\epsilon(n)$ ball around the destination as $\tau_2(n)$ in the first
scenario (uniform $\kappa$), and $\tau_3(n)$ in the adversarial
scenario. The following theorem provides bounds for the hitting time
in both these scenarios.

\begin{theorem}
\label{thm:quatheorem}

\emph{(i). Uniformly random $\kappa$:} Let $(L_2,\alpha_2)$ be the
polar representation of a point chosen from a semicircular sector
($\frac{-\pi}{2} < \alpha_2 < \frac{\pi}{2}$) of a unit circle, with a
probability distribution $\mathbf{f_{L_2,\alpha_2}}$ as in
(\ref{eqn:quaddistn}). Let $\beta_2=E(L_2\cos{\alpha_2})$.  Then, for
all positive $ c_5,c_6 : c_5 <\frac{1}{\beta_2} < c_6$, we have
$
\frac{c_5}{K}\sqrt{\frac{n}{\log(n)}} \le \tau_2(n) \le
\frac{c_6}{K}\sqrt{\frac{n}{\log(n)}} \quad(a.a.s).
$

\emph{(ii). Adversarial choice of quadrants:} Let $(L_3,\gamma)$ be
the polar representation of a point chosen randomly from a quadrant
containing the destination, where $\gamma$ is the angle with respect
to the local quadrant.  Let $\beta_3=E(L_3
\min(\cos{\gamma},\sin{\gamma}))$. Then for all positive $ c :
\frac{1}{\beta_3} < c$, we have $\tau_3(n) \le
\frac{c}{K}\sqrt{\frac{n}{\log(n)}} \quad(a.a.s).$

\end{theorem}
 \begin{proof}
 \emph{(i). Uniformly random $\kappa$:} Consider any node B in the
 semicircle AFC in Figure~\ref{fig:quad1dia}(b). For any step
 $i<\tau_2(n)$, we have the following bounds for $Z^{(n)}(i)$, the
 distance traveled towards the destination in the $i^{th}$ step.
 The following bounds are similar to equations (\ref{eqn:lowbou})
 and (\ref{eqn:upbou}).
 \begin{equation}
 Z^{(n)}(i) =  |\overrightarrow{OA}|- |\overrightarrow{OB}|,
 \end{equation}
 \begin{eqnarray}
 \hat{L}_{2}^{(n)}(i)\cos{\alpha^{(n)}_2(i)} -
 \frac{{\hat{L}_{2}^{(n)}(i)}^2}{\epsilon(n)}\le Z^{(n)}(i) \nonumber
 \\ \le \hat{L}_{2}^{(n)}(i)\cos{\alpha^{(n)}_2(i)}.
 \end{eqnarray}
 The rest of the proof is analogous to Theorem ~\ref{thm:sectheta},
 where we substitute $\hat{L}_{2}^{(n)}$ for $\hat{L}^{(n)}$,
 $\tau_2(n)$ for $\tau(n)$, $Z^{(n)}$ for $Y^{(n)}$, $\alpha_2$ for
 $\alpha$ and $\beta_2$ for $\beta$. The details are skipped for
 brevity.
 \emph{(ii). Adversarial choice of quadrants:} From
 Figure~\ref{fig:quad1dia}(a), it is clear that once a node `B' is
 selected, the distance traveled towards the destination is
 minimized if the destination O was along either O' or O'',
 whichever is more unfavorable. Thus, the distance traveled towards
 the destination in the $i^{th}$ step is bounded below by
 \begin{eqnarray}
 \min \Big\{ \hat{L}_{3}^{(n)}(i)\cos{\gamma^{(n)}(i)} -
 \frac{{\hat{L}_{3}^{(n)}(i)}^2}{\epsilon(n)},\nonumber
 \\\hat{L}_{3}^{(n)}(i)\sin{\gamma^{(n)}(i)} -
 \frac{{\hat{L}_{3}^{(n)}(i)}^2}{\epsilon(n)} \Big\} \le Z^{(n)}(i).
 \end{eqnarray}
 By arguments similar to Theorem \ref{thm:sectheta}, we can show
 that $\tau_3(n) \le c\frac{1}{M(n)} \quad(a.a.s).$ \end{proof}

The quadrant information can
be replaced by half-plane information and still lead to a delay that
is within a constant factor of straight-line routing, if the uniform
$\kappa$ assumption is made.  However, half-plane information is not
sufficient for order-wise straight-line routing delay in an
adversarial scenario.

\section{Routing with Fractional Information}
\label{sec:frac}

In this section, we consider the case where only a small fraction of
the nodes have any routing information at all. Most nodes simply
forward the packet to a randomly selected neighbor. A small fraction
of the nodes have routing information (either quadrant information, or
GPS information with errors). Such routing information could be
distributed by some gossip mechanism (routing table updates)
\cite{limama99,kermasgan01}, where nodes forward routing information,
but also could clear routing tables after some time. We do not
explicitly model the dynamics of such messaging. Instead, we adopt the
following simple model for routing.

We assume that each point has routing information (either imprecise
GPS, or quadrant information) with a fixed probability $p \in (0,1),$
independent of any other event. With probability $1-p,$ the next hop
location is uniformly chosen from a circle of radius $M(n)$ about the
current location (i.e., random routing). In this section, we
explicitly derive the results only for the quadrant routing strategy.
Analogous results hold when only a fraction of the nodes have
imprecise GPS information.  With such a strategy, let us denote the
event that the $i^{th}$ hop location contains quadrant
information\footnote{For notational convenience, we suppress
  explicitly showing the dependence of $E(i)$ on $n$} by $E(i).$ As
before, we normalize the distance between the source and destination,
and denote the routing delay under the strategy described above by the
random variable $\tau_p(n).$

\begin{theorem}
\label{thm:thmwithp} Let $\beta_2$ be defined as in Theorem
~\ref{thm:quatheorem}. Then for all positive $c_1,c_2 : c_1 <
\frac{1}{p\beta_2} < c_2$, we have
\begin{displaymath}
\frac{c_1}{K}\sqrt{\frac{n}{\log(n)}} \le \tau_p(n) \le
\frac{c_2}{K}\sqrt{\frac{n}{\log(n)}} \quad(a.a.s).
\end{displaymath}
\end{theorem}
 \begin{proof}
 Let $Q^{(n)}(i)$ be the random distance traveled towards the
 destination in the $i^{th}$ step. Then,\begin{equation} Q^{(n)}(i)
 = Z^{(n)}(i)1_{E(i)} +  R^{(n)}(i)1_{E^c(i)}
 \end{equation}
 where $Z^{(n)}(i)$ is the random distance traveled towards destination
 with a quadrant information strategy, and $R^{(n)}(i)$ is the distance
 traveled without any information. Let `B' be the next hop location if
 event $E(i)$ occurs, else let the next hop location be `B1'.  Let
 $(\hat{L}_{p}^{(n)}(i),\alpha^{(n)}_L(i))$ and
 $(\hat{S}_{p}^{(n)}(i),\alpha^{(n)}_S(i))$ be the polar coordinates of
 the nodes `B' and `B1' respectively. Location `B' is defined
 identically to the next hop location in Section~\ref{sec:quad}; and
 location `B1' is chosen uniformly from a circle of radius $M(n)$ about
 `A'. Let us define
 \begin{eqnarray}
 P^{(n)}(i) = \{\hat{L}_{p}^{(n)}(i)\cos{\alpha^{(n)}_L(i)} -
 \frac{{\hat{L}_{p}^{(n)}(i)}^2}{\epsilon(n)} \}1_{E(i)} + \nonumber \\
 \{\hat{S}_{p}^{(n)}(i)\cos{\alpha^{(n)}_S(i)}
 -\frac{{\hat{S}_{p}^{(n)}(i)}^2}{\epsilon(n)}\}1_{E^c(i)}
 \end{eqnarray}
 \begin{eqnarray}\label{eqn:fractsimilar}
 T^{(n)}(i) = \hat{L}_{p}^{(n)}(i)\cos{\alpha^{(n)}_L(i)} 1_{E(i)} +
 \nonumber \\
 \hat{S}_{p}^{(n)}(i)\cos{\alpha^{(n)}_S(i)}1_{E^c(i)}
 \end{eqnarray}
 From Lemma~\ref{lem:randpointbound}, we have the following bound
 for all $i < \tau_p(n).$
 \begin{equation}\label{eqn:fractineq}
 P^{(n)}(i) \le Q^{(n)}(i) \le T^{(n)}(i).
 \end{equation}
 Now, let us suppose that
 \begin{displaymath}
 \tau_p(n) \le c_2\frac{1}{M(n)} \quad(a.a.s.)
 \end{displaymath} is not true for some $c_2 > \frac{1}{p\beta_2}$.
 Then, there exists a subsequence $n_k, k = 1,2,\dots$ such that
 $\tau_p(n_k) > \frac{c_2}{M(n)}$. Now, along this subsequence,
 \begin{displaymath}
  \sum_{i=1}^{\frac{c_2}{M(n_k)}} P^{(n_k)}(i)
 =M(n_k)\sum_{i=1}^{\frac{c_2}{M(n_k)}}
 \{{L}_{p}^{(n_k)}(i)\cos{\alpha^{(n_k)}_L(i)}
 \end{displaymath}
 \begin{displaymath}
 -M(n_k)\frac{{{L}_{p}^{(n_k)}(i)}^2}{\epsilon(n_k)}\}1_{E(i)} +
 M(n_k)\sum_{i=1}^{\frac{c_2}{M(n_k)}}
 \{{S}_{p}^{(n_k)}(i)\cos{\alpha^{(n_k)}_S(i)}
 \end{displaymath}
 \begin{equation}\label{eqn:lobouwithP}
 -M(n_k)\frac{{{S}_{p}^{(n_k)}(i)}^2}{\epsilon(n_k)}\}1_{E^{c}(i)}
 \end{equation}
 The terms on the $R.H.S$ of  (\ref{eqn:lobouwithP}) are a
 triangular array of i.i.d. random variables. Noting that
 ${S}_{p}^{(n_k)}(i)\cos{\alpha^{(n)}_S(i)}$ is a symmetric random
 variable with mean zero, the limit of the sum in equation
 (\ref{eqn:lobouwithP})
 \begin{equation}
 \lim_{n\rightarrow\infty} \sum_{i=1}^{\frac{c_2}{M(n_k)}}
 P^{(n_k)}(i)= c_2\beta p
 \end{equation}which is greater than unity. This contradicts the fact
 that
 $\frac{c_2}{M(n)}$ was smaller than $\tau_p(n)$, as the path has
 already reached the destination. Hence,
 \begin{displaymath}
 \tau_p(n) \le \frac{c_2}{M(n)} \quad \forall c_2 >
 \frac{1}{p\beta}\quad (a.a.s.).
 \end{displaymath}
 Similarly, using equation(\ref{eqn:fractsimilar}) we show that
 \begin{displaymath}
 \tau_p(n) \ge \frac{c_1}{M(n)} \quad \forall c_1 <
 \frac{1}{p\beta}\quad (a.a.s).
 \end{displaymath} \end{proof}

\section{Simulation Results for Routing Delay}
\label{sec:sim}

We have so far assumed a continuum model of nodes in the unit
square. In this section, we account for the discretization effects,
and simulate the various scenarios discussed earlier.  We consider a
simulation scenario where $N$ nodes are placed uniformly randomly on a
unit square. The source is located at $[0,0]$ and the destination at
$[0.7,0.7]$ (such that the Euclidean distance between the source and
the destination is one). A histogram of the routing delays (number of
hops) from 150 simulations is plotted, along with a sample path for
illustration.

\begin{figure}[h]
\begin{center}
\begin{tabular}{cc}
\includegraphics*[width=0.43\columnwidth]{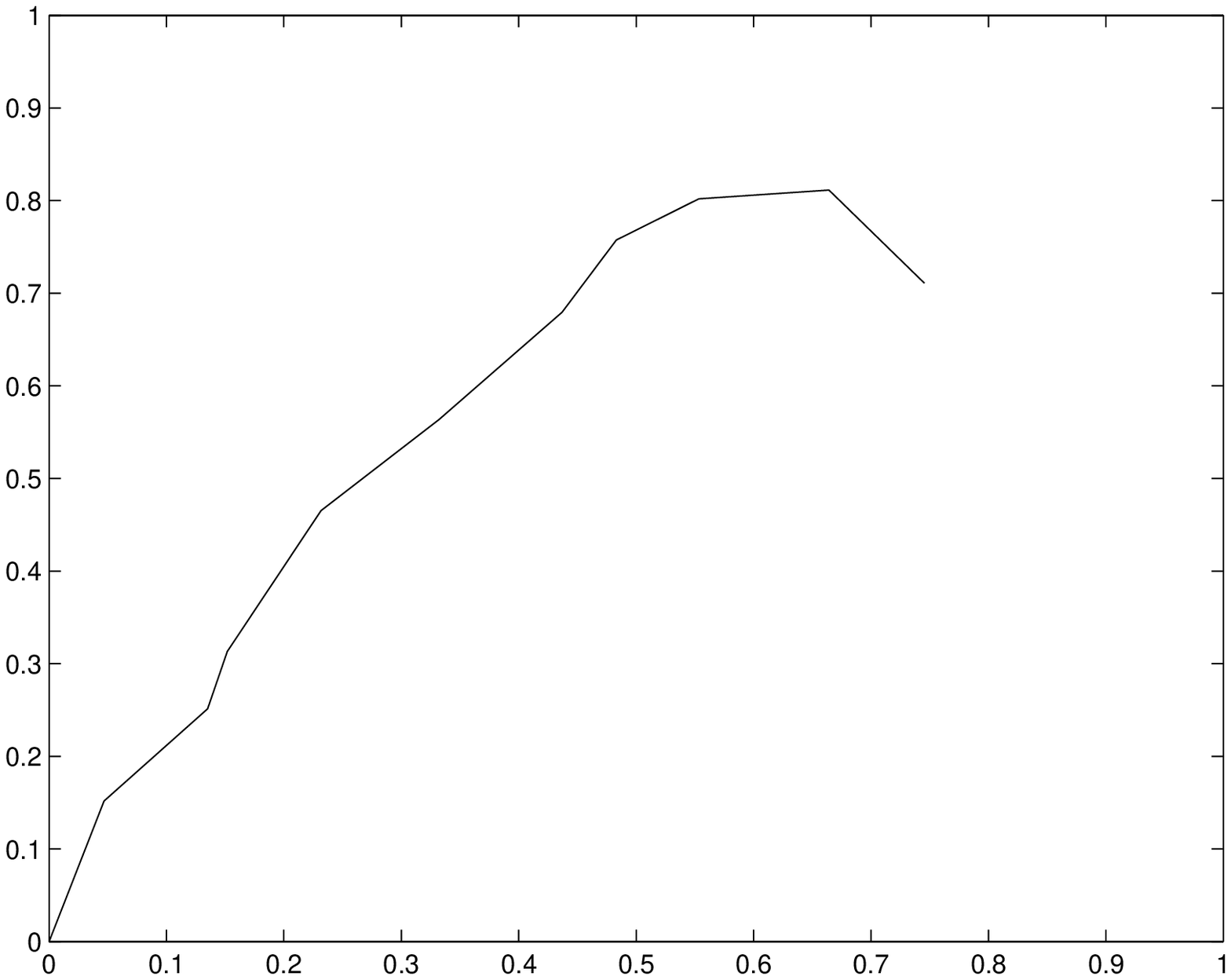}&
\includegraphics*[width=0.43\columnwidth]{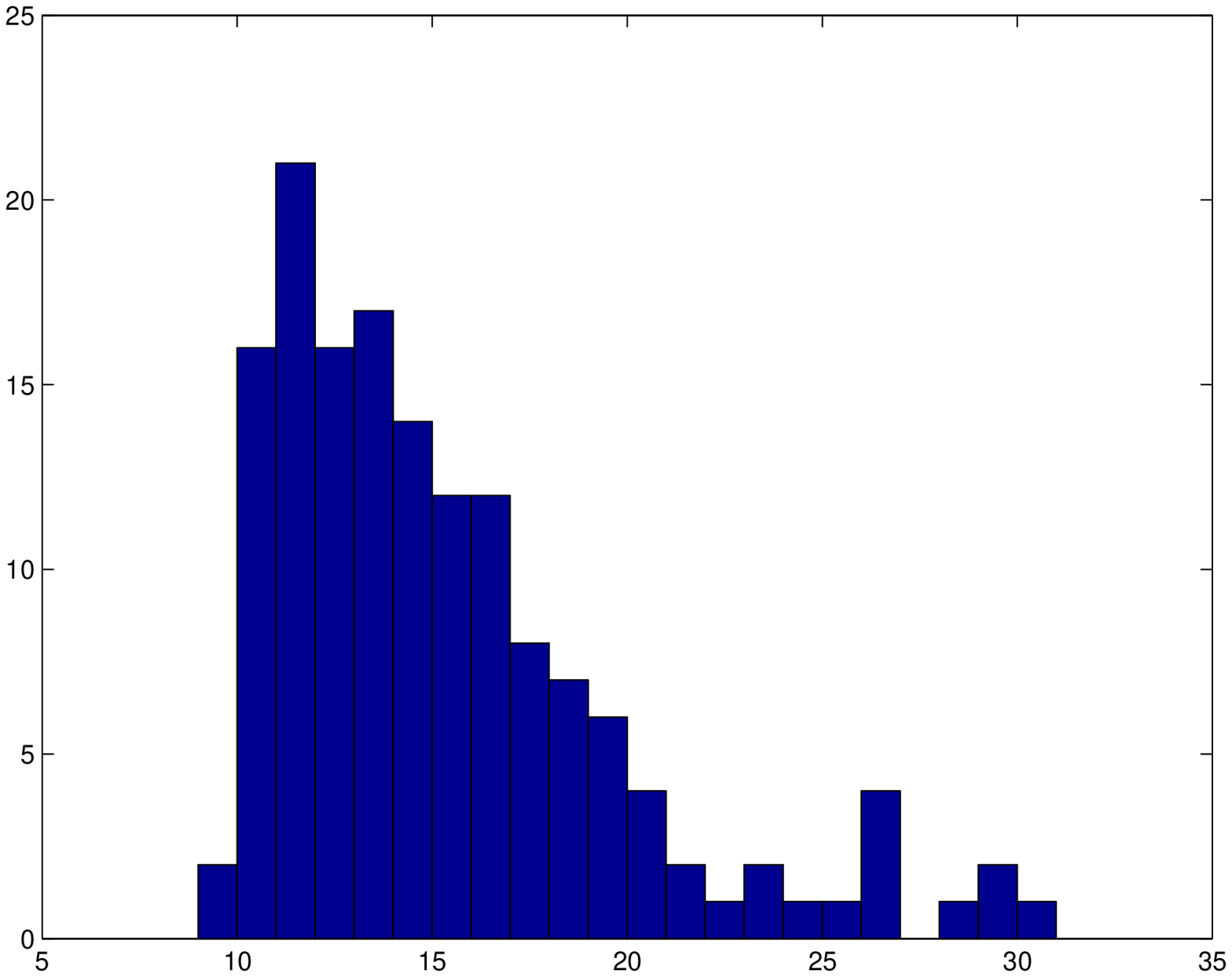}
\\ (a) Sample Path  & (b) Distribution
\end{tabular}
\caption{Biased Sector Routing- Spiraling drift - 1000
nodes\label{fig:seczer451}}
\end{center}
\end{figure}

\begin{figure}[h]
\begin{center}
\begin{tabular}{cc}
\includegraphics*[width=0.43\columnwidth]{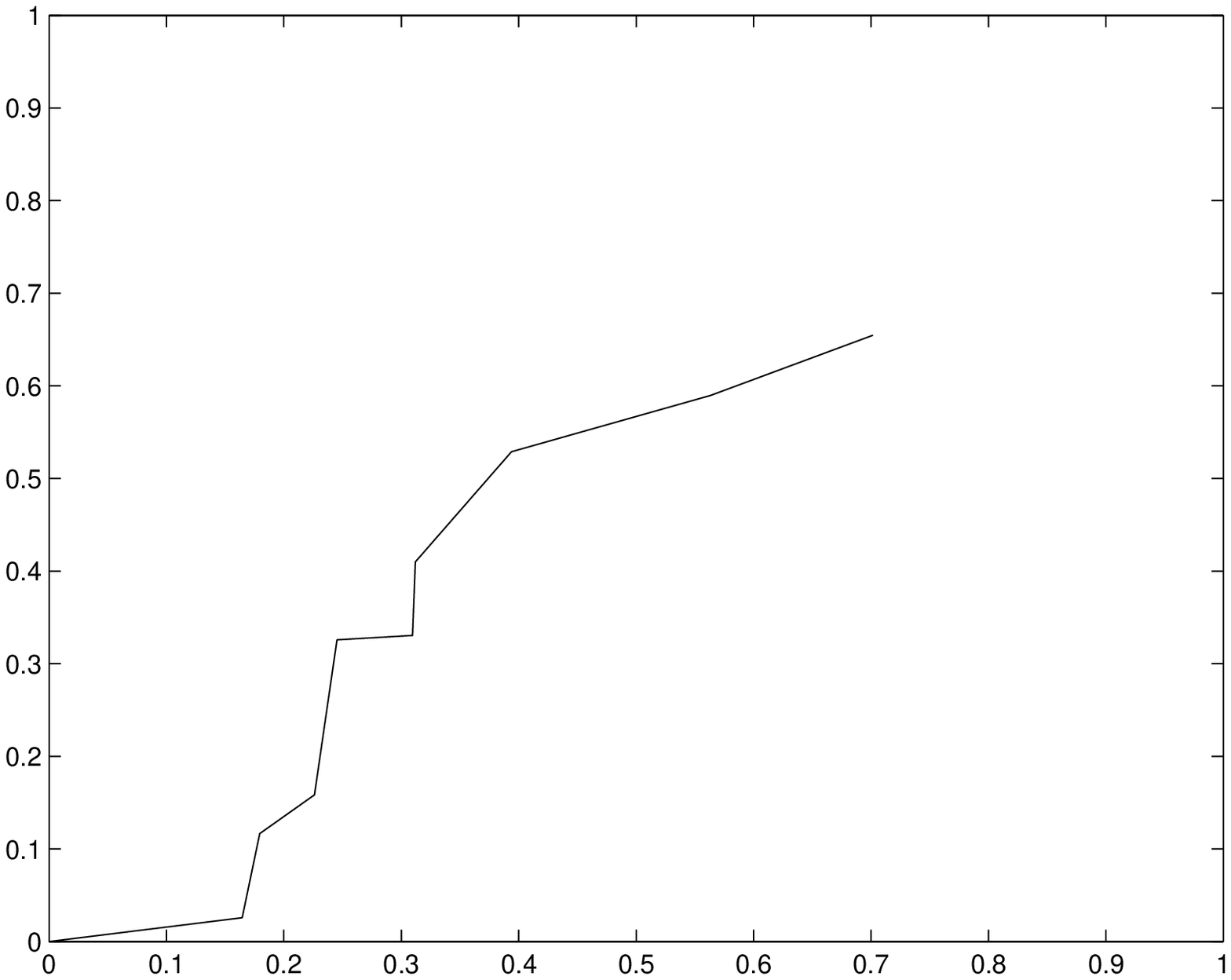}&
\includegraphics*[width=0.43\columnwidth]{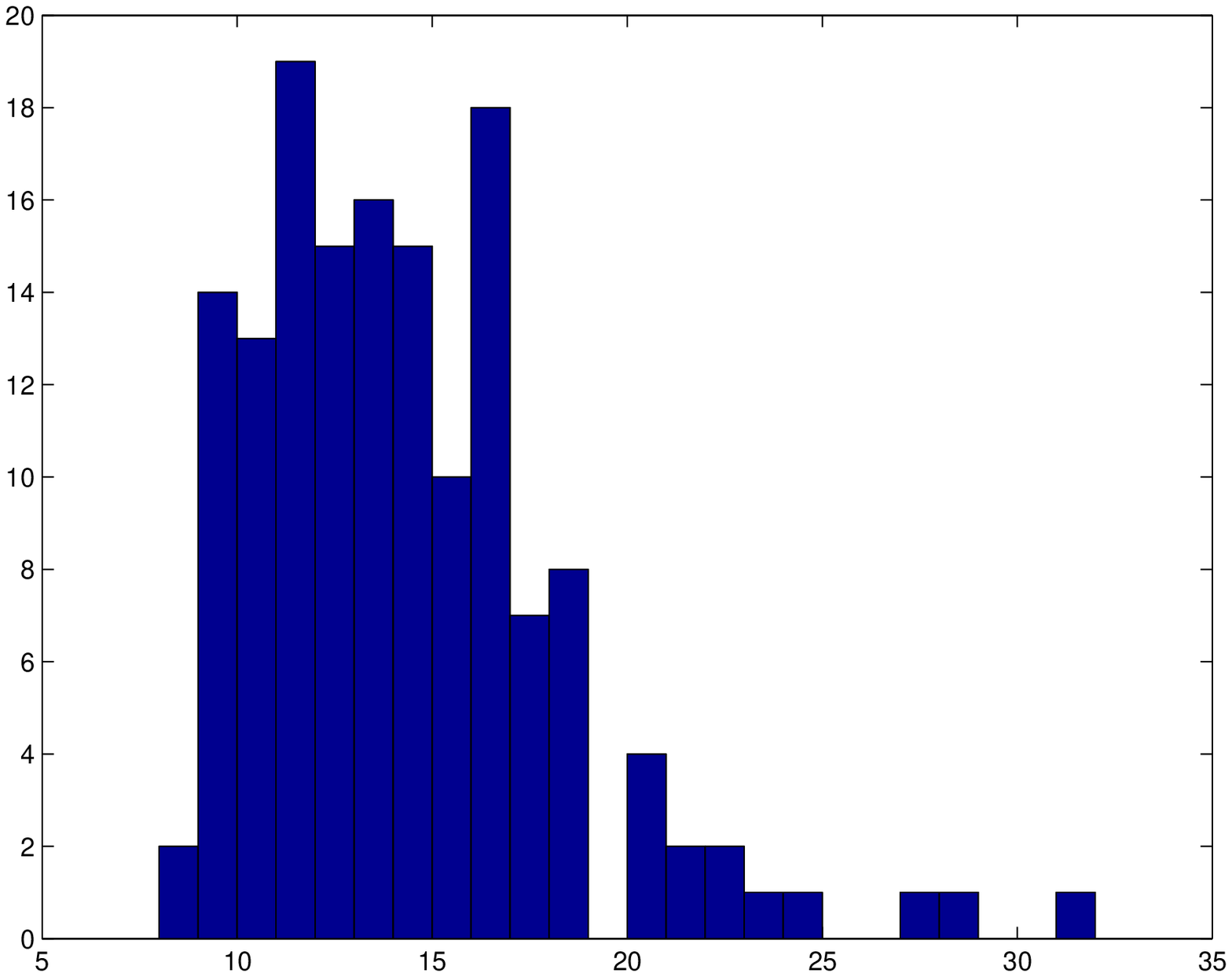} \\
(a) Sample Path  & (b) Distribution
\end{tabular}
\caption{Quadrant based Routing - 1000 nodes\label{fig:quad1}}
\end{center}
\end{figure}

\begin{figure}[h]
\begin{center}
\begin{tabular}{cc}
\includegraphics*[width=0.43\columnwidth]{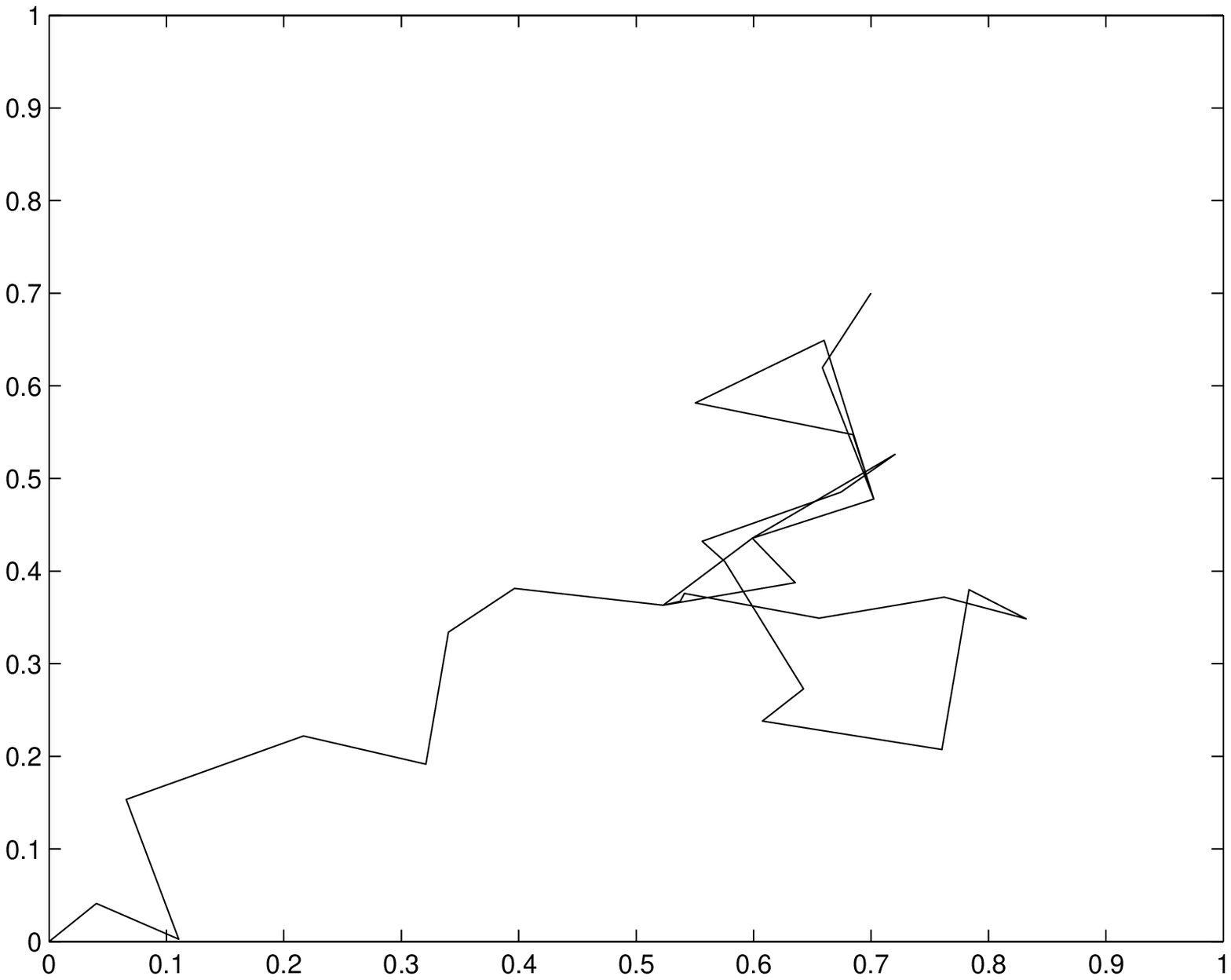}&
\includegraphics*[width=0.43\columnwidth]{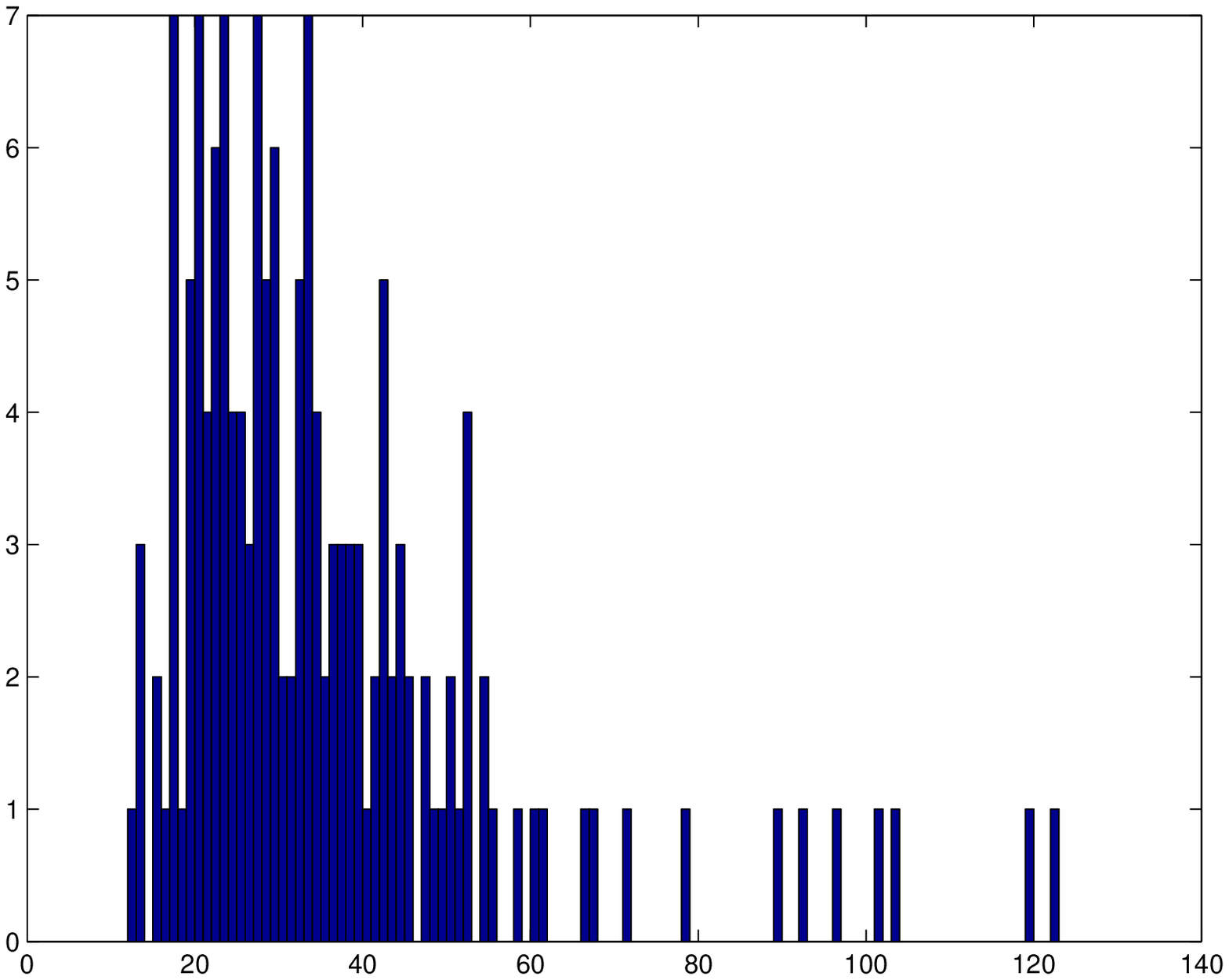} \\
(a) Sample Path  & (b) Distribution
\end{tabular}
\caption{Fractional information: 35\% have quadrant information
-1000 nodes \label{fig:frac1}}
\end{center}
\end{figure}

The simulations are for a node density $N=1000$. 
The geographic greedy routing strategy, where
the relay node is the neighbor node that is closest to the destination
shows an almost deterministic path length of 7 hops and the
corresponding sample path resembles a straight-line path from the
source to the destination. 
This simulation corresponds to a
``small-scale'' network as the number of hops with straight-line
routing is relatively small. 
The small variations in the path length occur due to the randomness in
the node positions.  With unbiased sectors (of 60 degrees), our
simulation results indicate that the average path length is about 11
hops, which is an increase by a factor predicted in
Theorem~\ref{thm:sectheta} (the analysis predicts a length of 11.01).
Routing with biased GPS
information is considered next, and the sample path shows some
spiraling (Figure~\ref{fig:seczer451}(a)) due to bias in routing,
and the average routing delay is about 15 hops. The quadrant based
routing strategy is simulated next in this setup, and the results are
shown in Figure~\ref{fig:quad1}.  The sample path is observed to be
similar to the sector routing case, and the average routing delay of
15 hops is only marginally more than the sector routing strategy.
Both of these are again close to that predicted by our analytical
results. Routing with fractional information is simulated by assuming
that a node contains routing (quadrant) information with a probability
of $p=.35.$ The sample path and the distribution of routing delay are
shown in Figure~\ref{fig:frac1}. The routing path is considerably
lengthened as most of the nodes do not contain routing information.
The average delay in this case is approximately 40 hops, which is
close to the analytically predicted value (42.7 hops). These plots
indicate that the random routing strategies have delays that are
comparable to the greedy geographic routing strategy, as predicted by
our analysis.

\begin{figure}[t!]
 \begin{center}
 \begin{tabular}{cc}
 \includegraphics*[width=0.43\columnwidth]{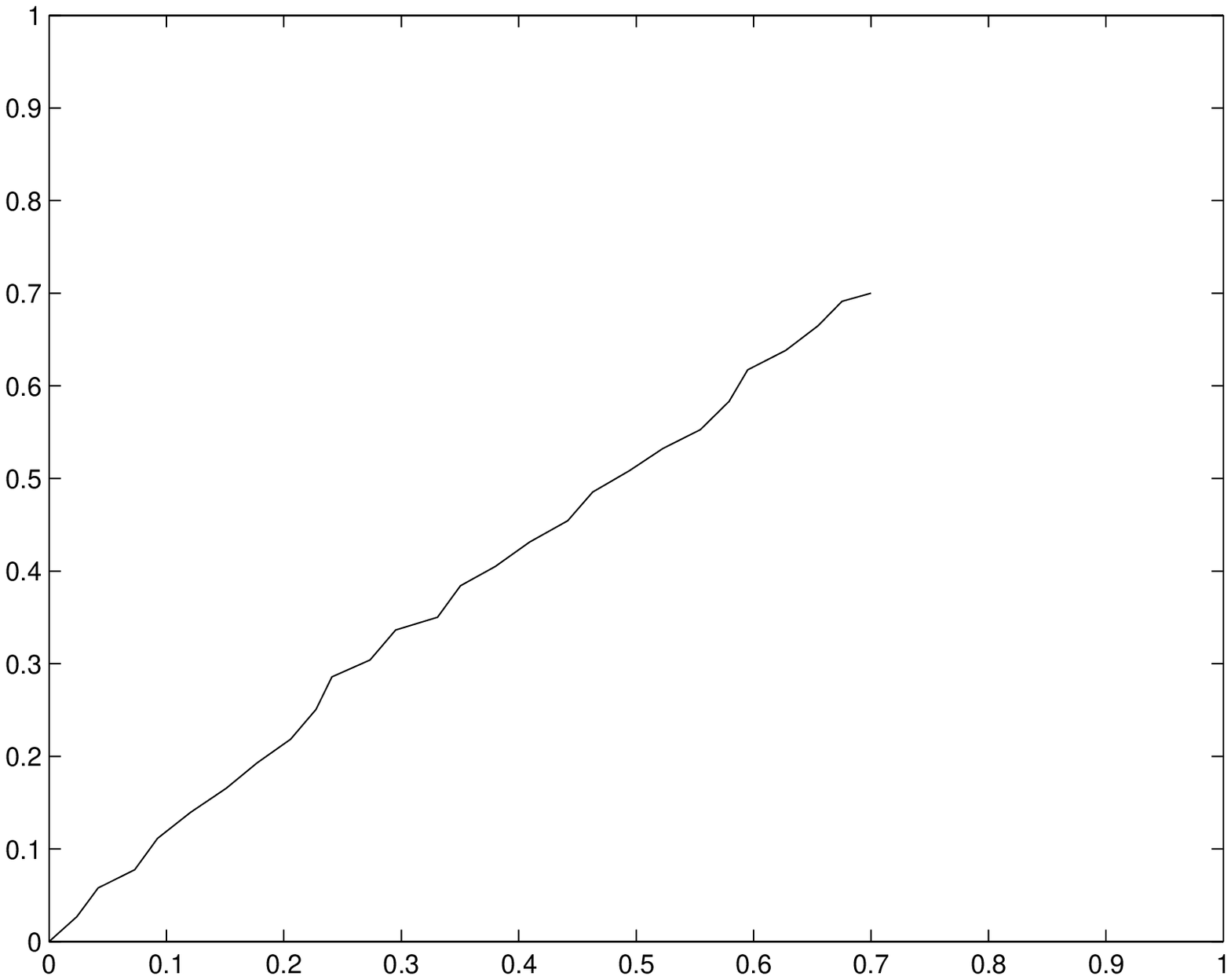}&
 \includegraphics*[width=0.43\columnwidth]{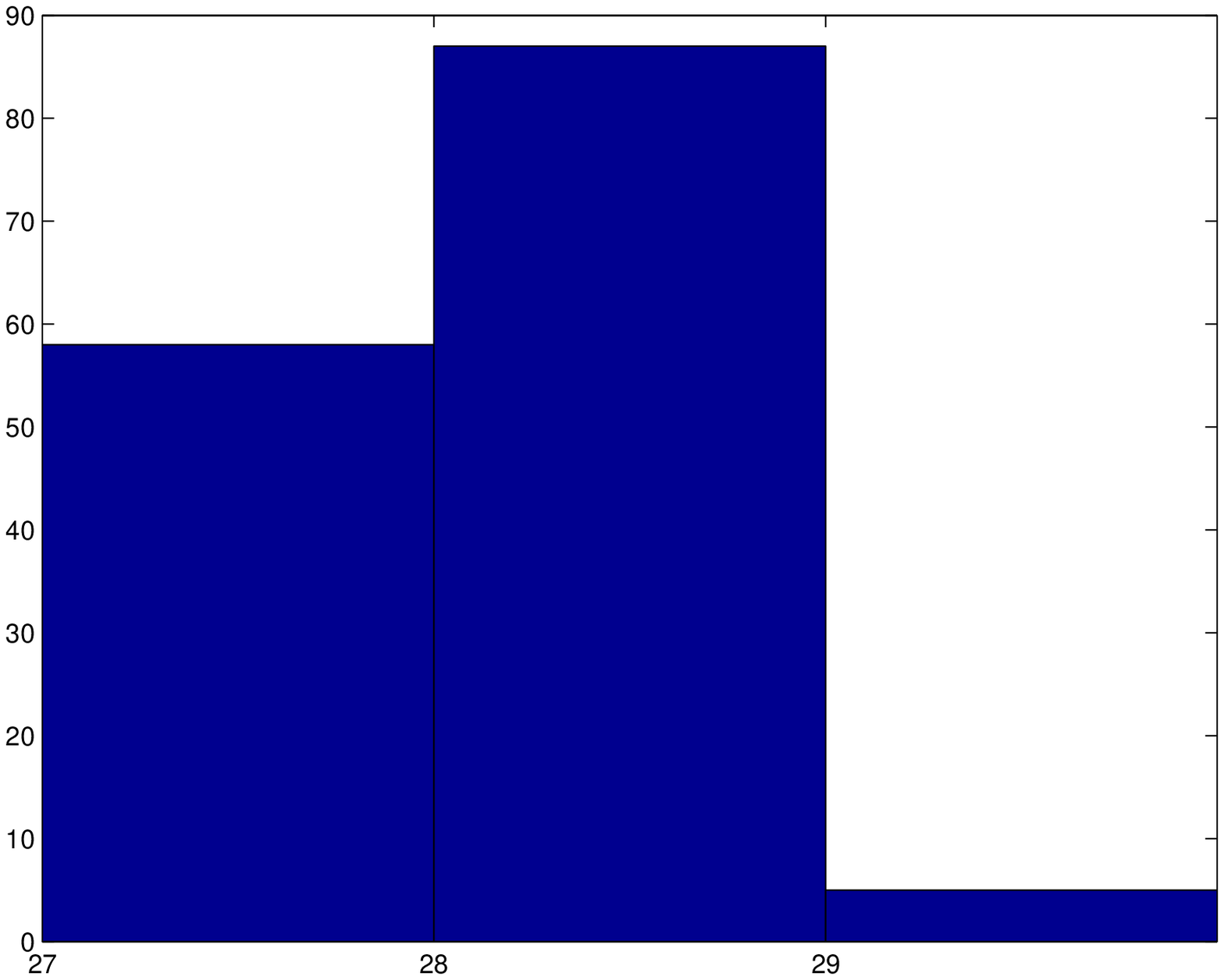} \\
 (a) Sample Path  & (b) Distribution
 \end{tabular}
 \caption{Straight Line Routing - 10,000 nodes \label{fig:stl10}}
 \end{center}
 \end{figure}

\begin{figure}[h]
 \begin{center}
 \begin{tabular}{cc}
 \includegraphics*[width=0.43\columnwidth]{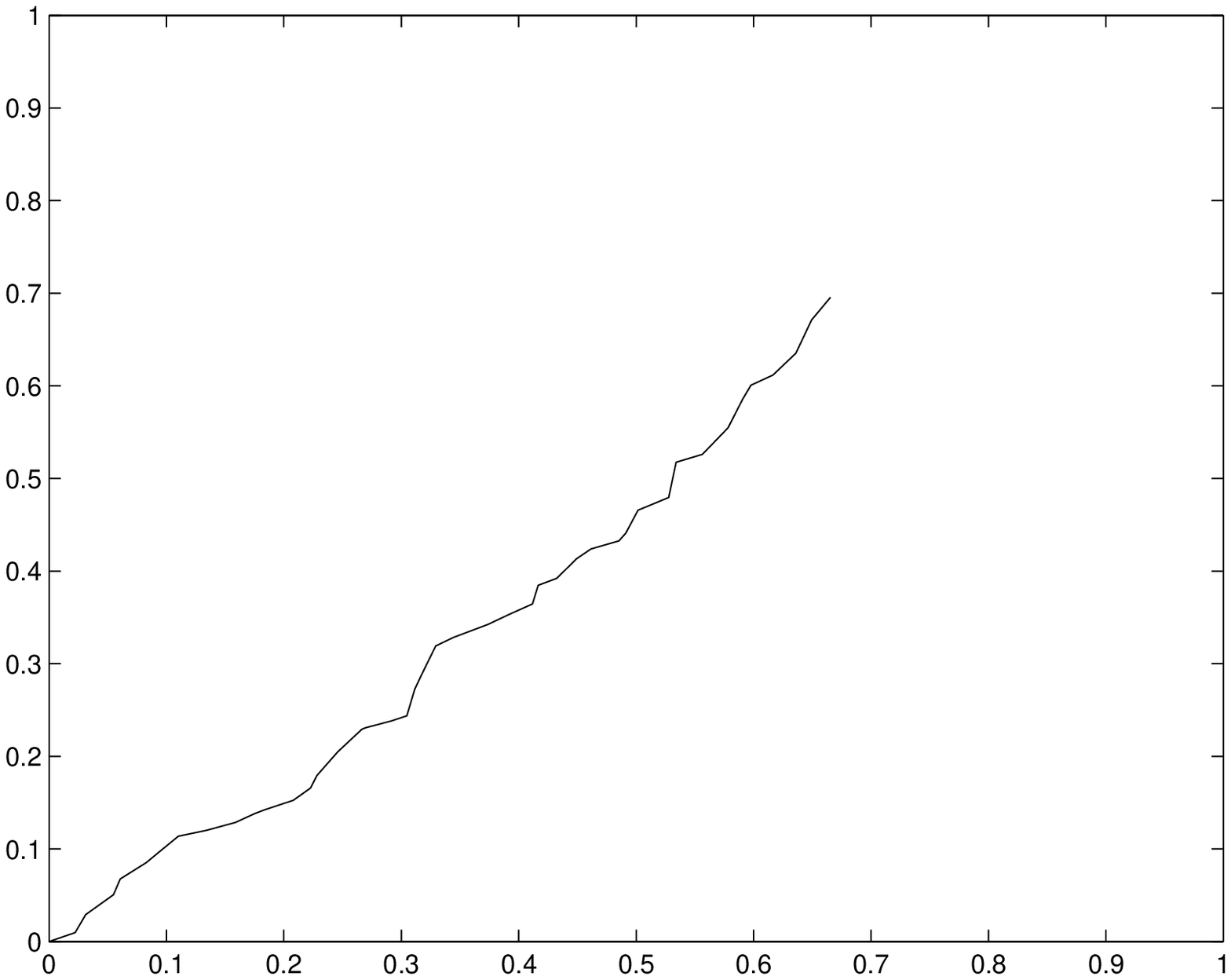}&
 \includegraphics*[width=0.43\columnwidth]{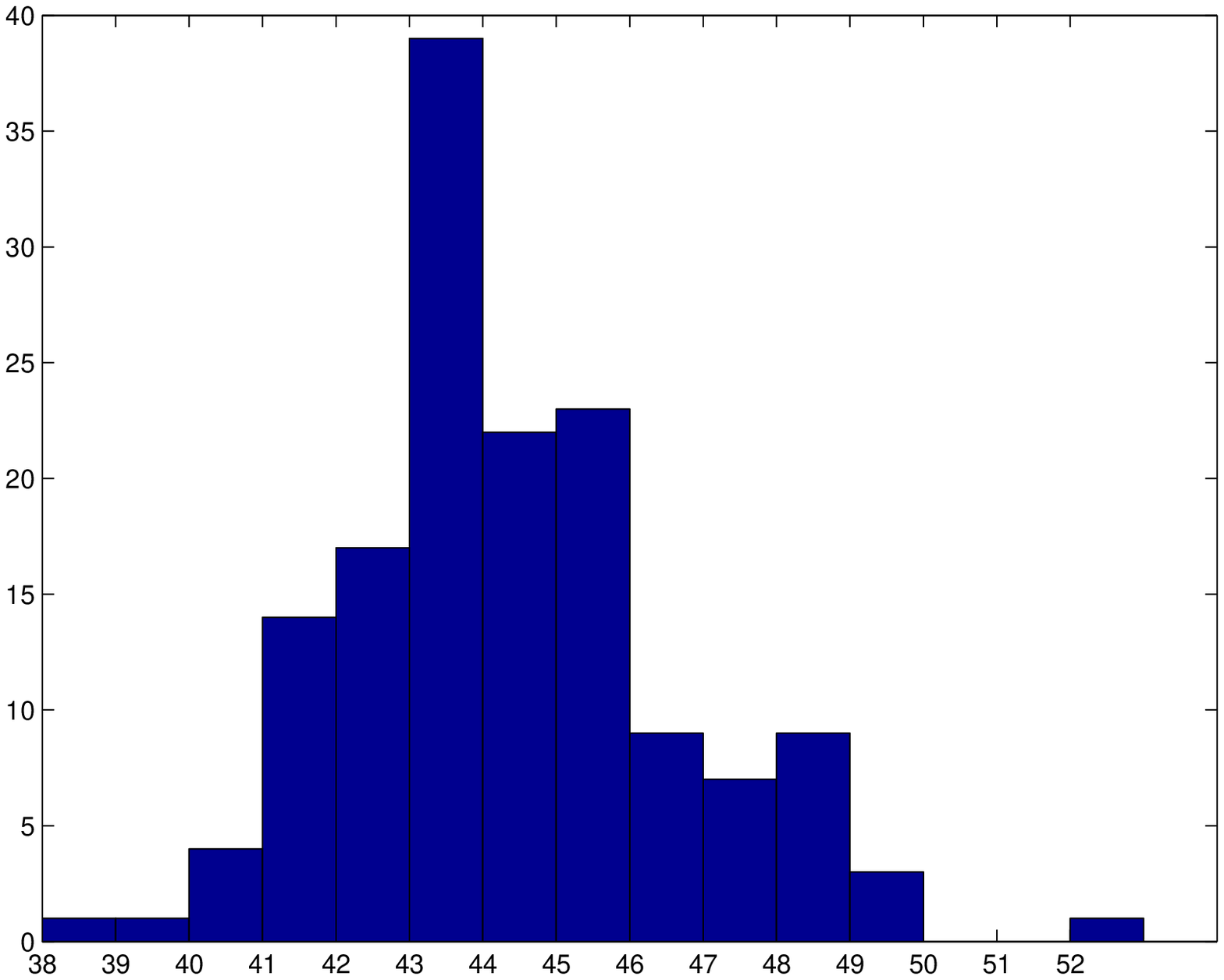} \\
 (a) Sample Path  & (b) Distribution
 \end{tabular}
 \caption{Unbiased Sector Routing - 10,000 nodes\label{fig:sec10}}
 \end{center}
 \end{figure}

\begin{figure}[h]
 \begin{center}
 \begin{tabular}{cc}
 \includegraphics*[width=0.43\columnwidth]{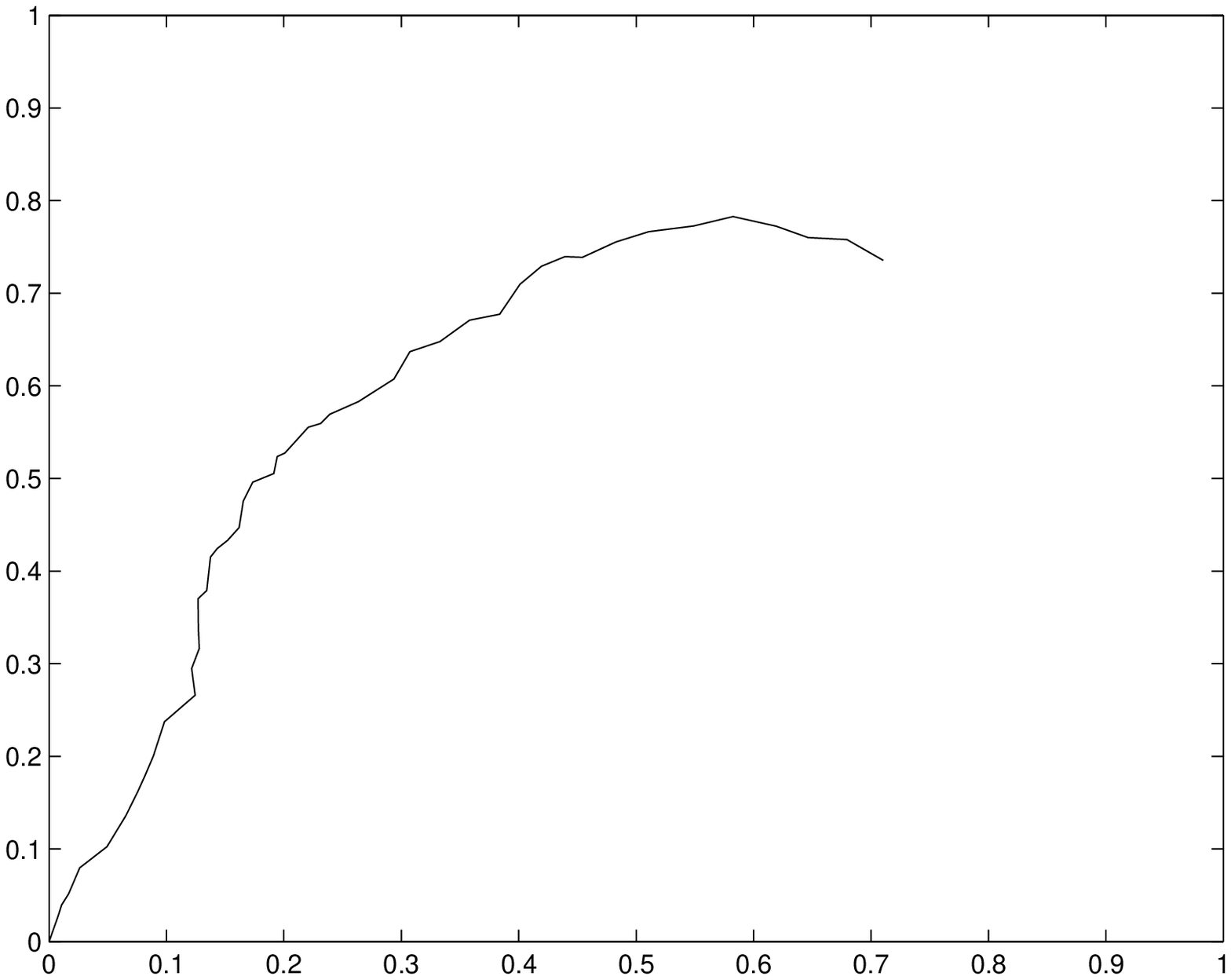}&
 \includegraphics*[width=0.43\columnwidth]{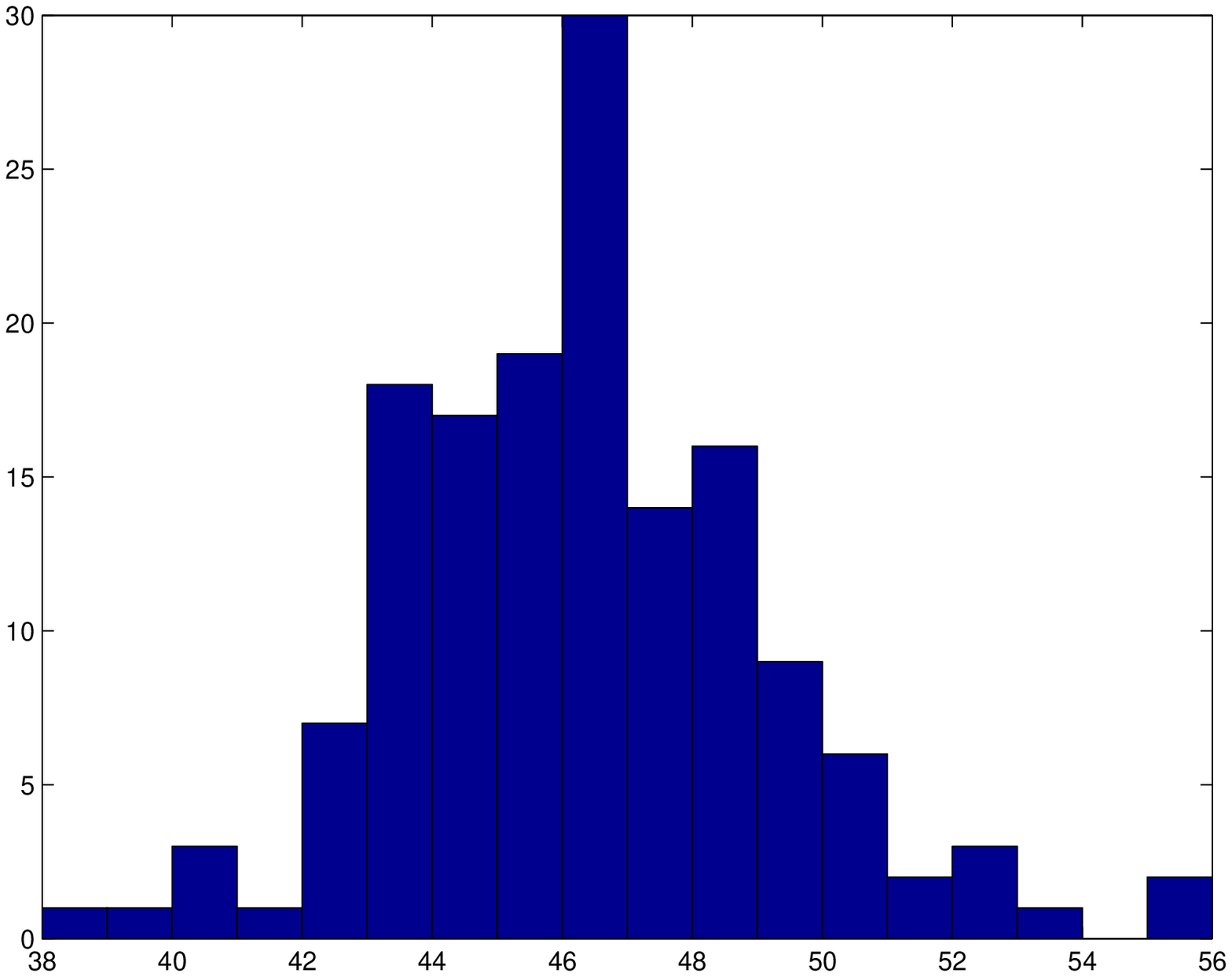} \\
 (a) Sample Path  & (b) Distribution
 \end{tabular}
 \caption{Biased Sector Routing- Spiralling drift - 10,000
 nodes\label{fig:seczer4510}}
 \end{center}
 \end{figure}

\begin{figure}[h]
 \begin{center}
 \begin{tabular}{cc}
 \includegraphics*[width=0.43\columnwidth]{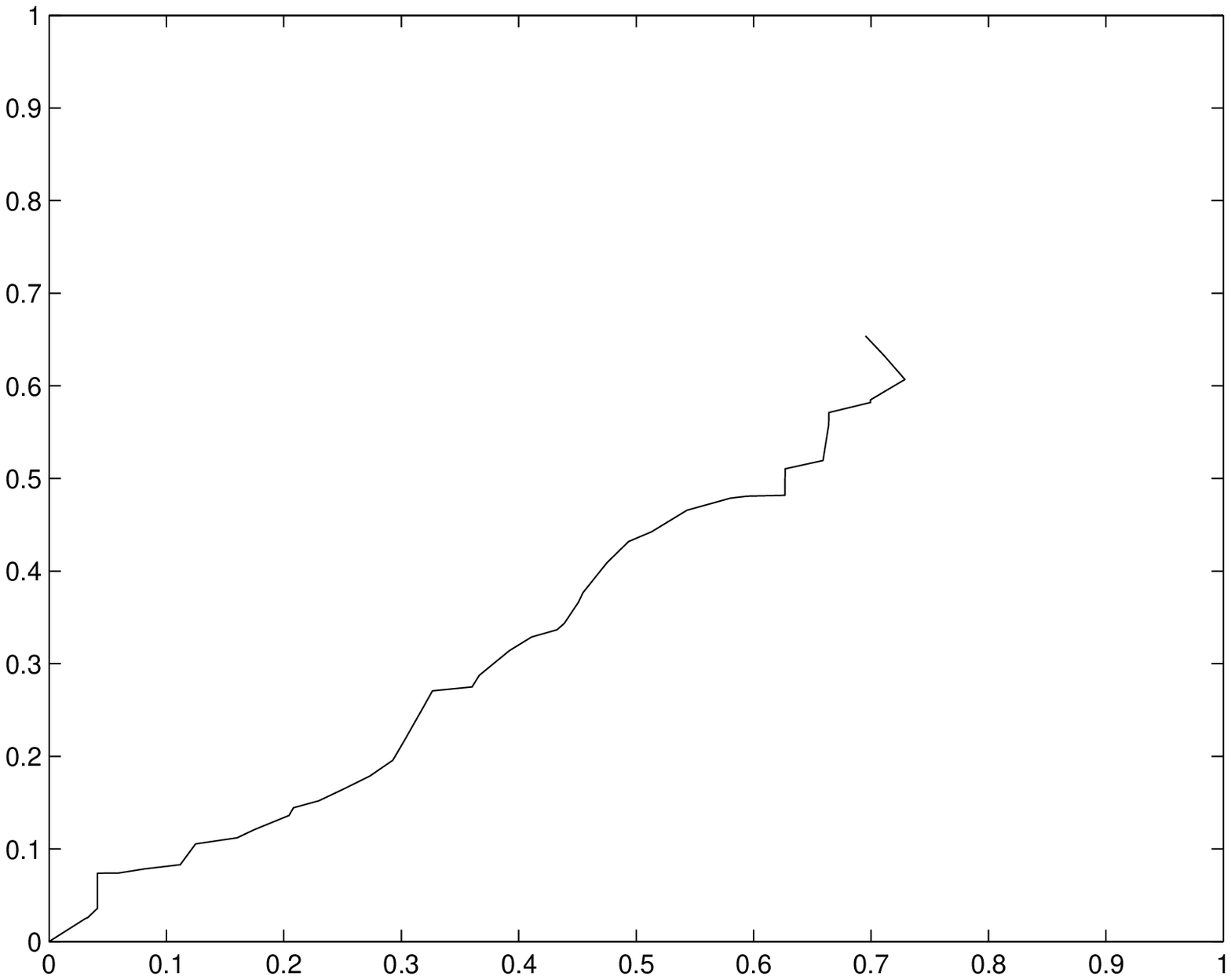}&
 \includegraphics*[width=0.43\columnwidth]{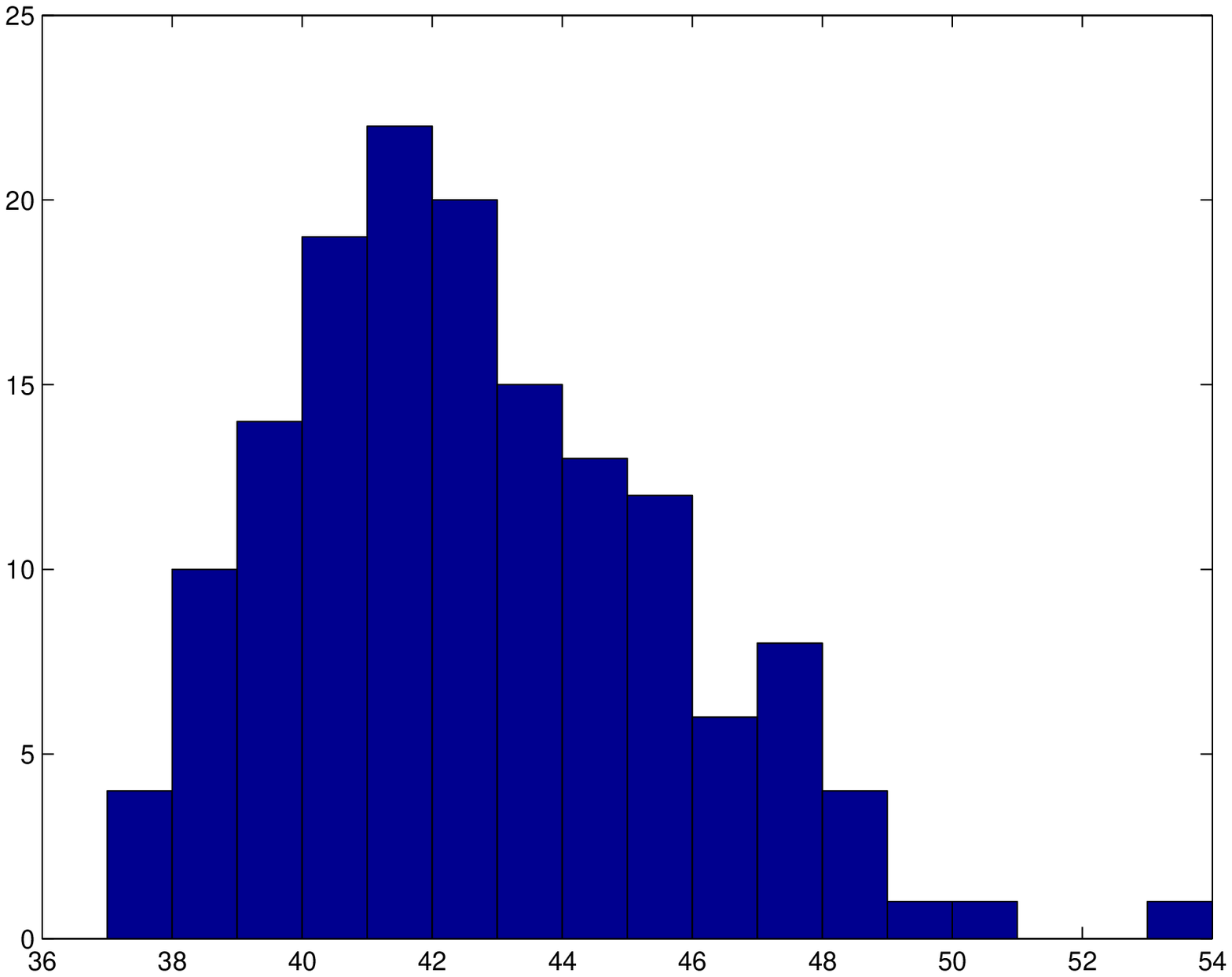} \\
 (a) Sample Path  & (b) Distribution
 \end{tabular}
 \caption{Quadrant based Routing - 10,000 nodes\label{fig:quad10}}
 \end{center}
 \end{figure}

\begin{figure}[h]
 \begin{center}
 \begin{tabular}{cc}
 \includegraphics*[width=0.43\columnwidth]{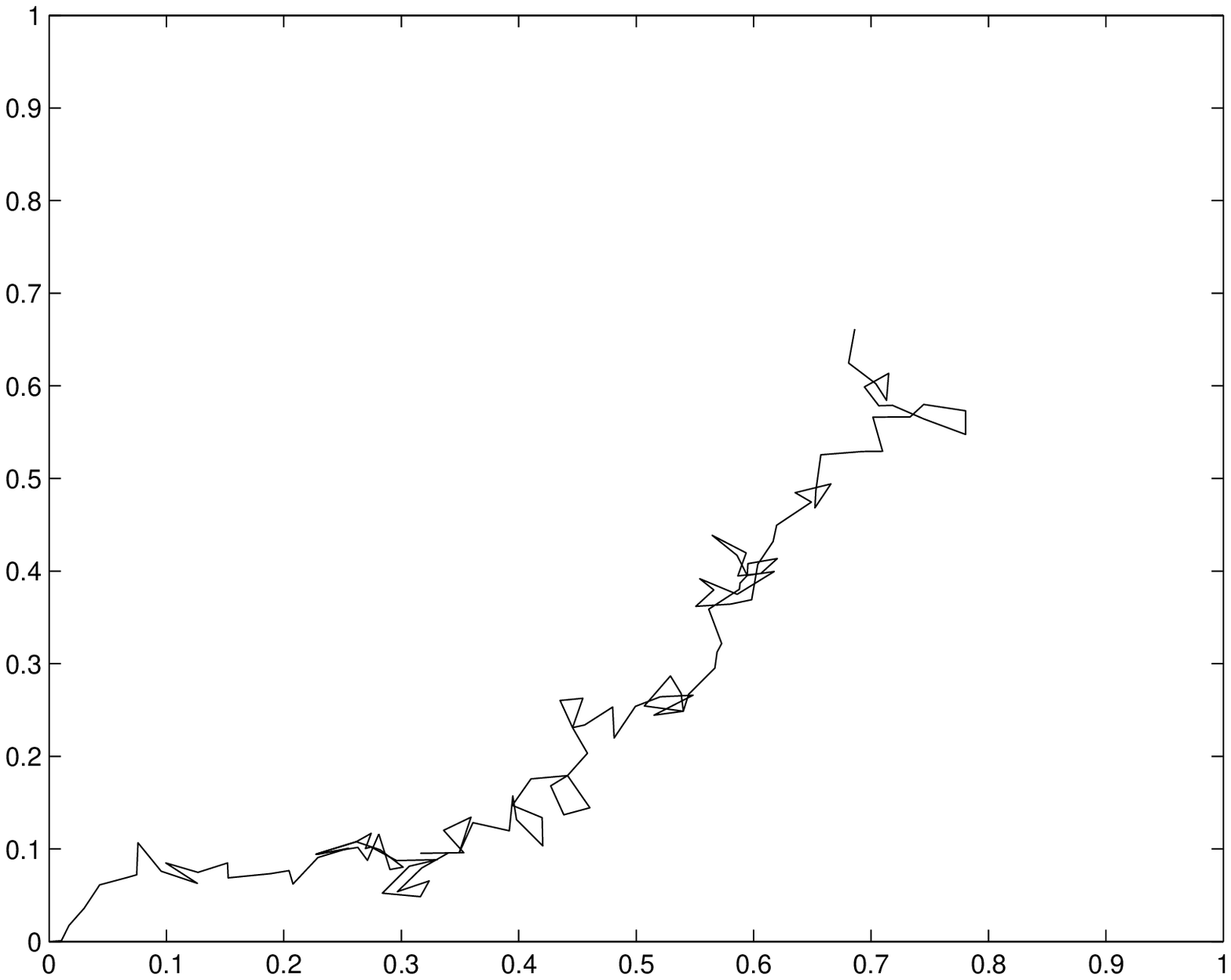}&
 \includegraphics*[width=0.43\columnwidth]{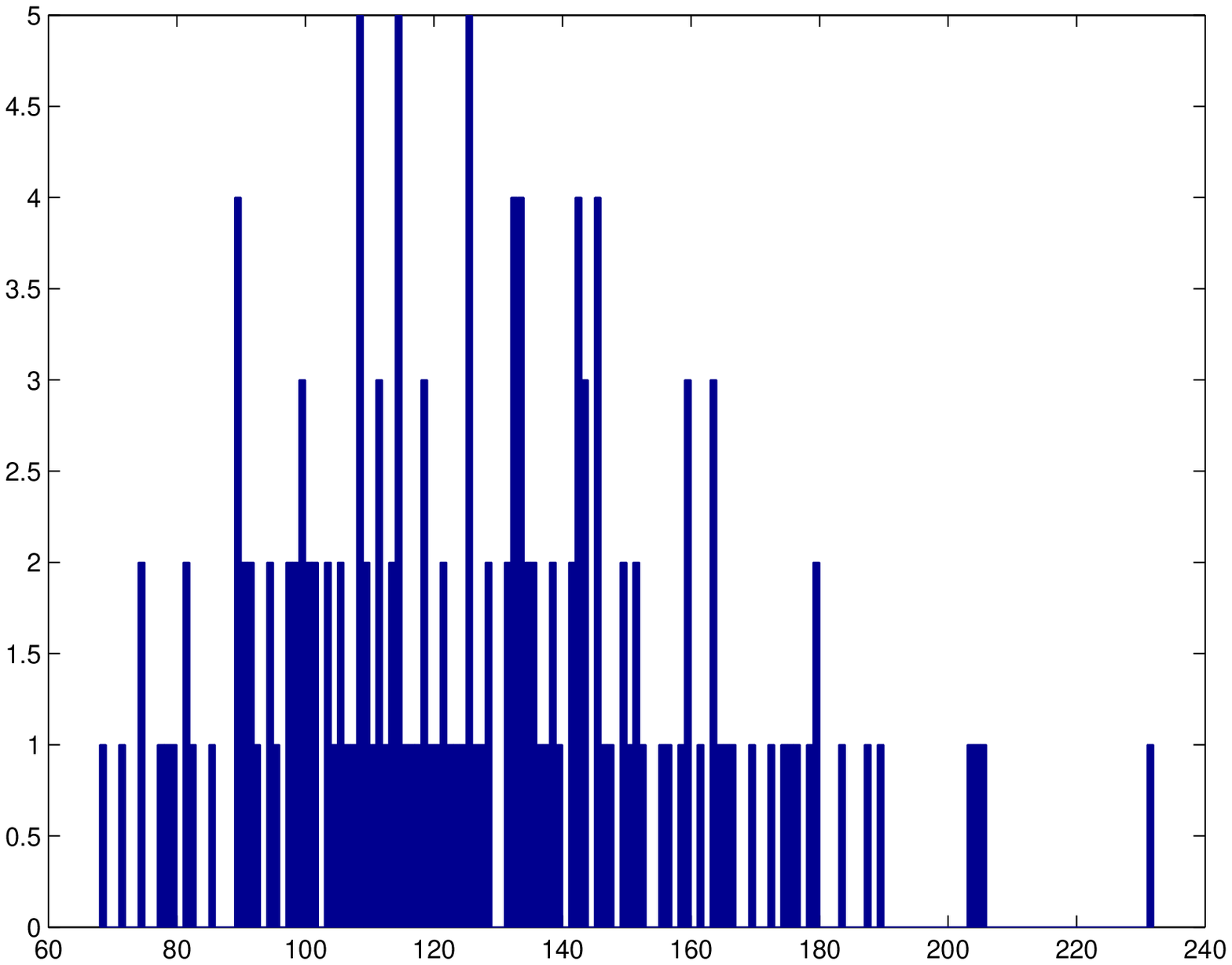}
 \\
 (a) Sample Path  & (b) Distribution
 \end{tabular}
 \caption{Fractional information: 35\% have quadrant information
 -10,000 nodes \label{fig:frac10}}
 \end{center}
 \end{figure}

The simulations are repeated for a larger network with $N=10000$
nodes. The number of hops for a greedy geographic routing strategy is
about 28 hops, which is about four times as that in the previous case.
The analogous results for the five routing strategies are displayed in
Figures~\ref{fig:stl10}--\ref{fig:frac10}. The average fractional
routing delay is about 120 hops (Figure~\ref{fig:frac10}(b)), which is
approximately a $\frac{1}{p}$ factor increase from the routing delay
for the quadrant routing scheme (which has an average routing delay of
42 hops). The spiralling drift of a routing scheme with directional
bias is also seen in Figure~\ref{fig:seczer4510}.

\section{Throughput Capacity with Progressive Routing}
\label{sec:cap} In the previous sections, we had assumed a
continuum model of a sensor network for the analysis of routing
delays. To obtain the throughput capacity, we need to consider
individual nodes and their data-rates. Thus, in this section, we use a
discrete node model of the sensor network. We assume that the $n$
nodes are randomly placed on a unit square, and as before, the
transmission radio range of the nodes is $M(n) =
K\sqrt{\frac{\log{n}}{n}}$. To avoid technical complications due to
edge effects, we assume that paths wrap around the edges of the unit
square.  Thus, the distance between any two nodes is simply the
shortest straight-line path between them (possibly with wrap-around).
The scaling parameters are the same as in the continuum model. We also
assume the Protocol Model \cite{gupkum00} for successful
transmissions.

\begin{definition} \label{def:promod} The transmission protocol is
called the \textit{protocol model} if the transmission from node
$A$ to $B$ is successful if $d(A,B) < M(n)$ and
$d(C_l,B) \ge (1+\Delta)M(n)$ for all other transmitting nodes
$C_l$.
\end{definition}

We consider \textit{progressive} routing strategies that ensure that
at each step of the route, the distance to the destination decreases
by at least $\delta M(n)$ for some $\delta >0.$ For example, routing
with sector information (considered in Section~\ref{sec:nobias}) will
lead to progressive routing if the transmit power exceeds a minimum
threshold, and the bias is not large.  We assume that a routing
strategy that satisfies this property is used for route setup, and
subsequent packets in each flow (between a source-destination pair)
follows this initial path. Further, the routes are independently setup
(across flows). It has been shown in \cite{gupkum00} that for routing
with straight-lines, the throughput capacity is
$\Theta{(\frac{1}{\sqrt{n\log{n}}})}$, for the protocol model, and the
upper bound on the throughput capacity with the protocol model is also
of this order.  However, with the addition of randomness in routing,
the capacity of the network could be reduced, as discussed in
Section~\ref{sec:int}.  The issue of concern is that the longer
routing paths due to the random strategies might create local hot
spots (see Figure~\ref{fig:routconf}(a)). We show that, for
progressive routing schemes, such local hot spots do not affect the
throughput capacity in an order-wise sense.

\begin{theorem}
  Consider a unit square, with $n$ nodes uniformly distributed, and
  $n/2$ randomly chosen source-destination pairs.  Let $\Psi$ be a
  progressive routing strategy such that in each hop, the Euclidean
  distance to the destination is reduced by $\delta M(n).$ Then, under
  the Protocol Model, a data rate of
  $\Theta{(\frac{1}{\sqrt{n\log{n}}})}$ is simultaneously achievable
  by every source-transmitter pair with routing strategy $\Psi$.
\end{theorem}
\begin{proof} We provide a sketch of the proof.
  Consider a uniform tiling of the unit square, by tiles of side
  $a(n) = \sqrt{\frac{\log{n}}{n}}$.  The idea of the proof is as
  follows:

\begin{enumerate}

\item We show that each tile is active (i.e., nodes in the tile are
    allowed to transmit) for a fixed fraction of the time, without
    being interfered by transmissions from other tiles.

  \item Observe that with progressive routing, each route could have
    multiple hops in each tile. We prove that an uniform upper bound
    on the number of hops in any tile summed over all routes is
    $\Theta{(\sqrt{n\log{n}})}$.

  \item Using these results, we show that each route receives a
    data-rate of $\Theta{(\frac{1}{\sqrt{n\log{n}}})}.$
\end{enumerate}
The above statements are proved in the following claims.

 It is clear from Definition~\ref{def:promod} that if there is a
 transmission from a node $A_i$ in some tile, other transmissions
 in neighboring tiles can affect the transmissions of $A_i$.
 However, since $\Delta$ is a constant, the number $J$ of nearby
 tiles that can affect the transmission is finite. We use this
 fact to construct a transmission schedule that allows for
 concurrent spatial transmissions. The problem is equivalent to a graph
 coloring problem with each vertex having at most a degree of $J$.
 Standard results from graph theory indicate that a graph with a
 degree no more than $J$ can have all its vertices colored by $J+1$
 colors such that no two neighbors have the same color. Thus, we
 can color the tiles with $J+1$ colors such that no two interfering
 tiles have the same color. We can construct a schedule such that a
 given slot is divided into $J+1$ sub-slots and all tiles of the
 same color can successfully transmit simultaneously.

 We assume that the strategy allows us to travel a distance
 of at least $\delta M(n)$ towards the destination in each jump.
 Thus, the number of hops required to reach the destination (that is
 a unit distance away) for any
 route is no more than $\frac{1}{\delta M(n)}$ hops. Also, since
 the maximum length that can be traveled in any hop is $M(n)$ the
 length of any routing path is upper bounded by $\frac{1}{\delta}$,
 an order $1$ quantity.

 \begin{claim}
 Given that a routing path passed through a tile, the number of
 hops inside the tile is no more than $\frac{\sqrt{2}}{\delta}$.
 \end{claim}
 \begin{proof}
 Assume that the required destination is outside the tile. Then,
 in $\frac{\sqrt{2}}{\delta}$ steps, the packet would have reached closer
 to the destination by more than $\sqrt{2} M(n)$, which would imply
 that the packet is no longer in the same tile. Even if all the
 intermediary steps fell inside the tile, the number of hops
 cannot be greater than $\frac{\sqrt{2}}{\delta}$. If the
 destination was inside the tile, it would have reach the
 destination within $\frac{\sqrt{2}}{\delta}$ steps.
 \end{proof}

 \begin{claim}
 The total number of tiles any path can touch is upper bounded by
 $\frac{1}{\delta M(n)}$.
 \end{claim}
 Since the total number of hops in any path is at most
 $\frac{1}{\delta M(n)}$ and all these hops can at best be in separate tiles,
 the claim holds.

 Let $X_{i}^{k}$ be a Bernoulli random variable, with $X_{i}^{k} =
 1$ if the $i^{th}$ path touched the $k^{th}$ tile. Clearly,
 $X_{i}^{k}$ is independent of $ X_{j}^{l} $ if $i \ne j$, as the paths are
 independently routed with respect to each other, although
 $X_{i}^{k} $ and $X_{i}^{l} $ are correlated. The random
 variable  $X_{i}^{k}$ is stochastically dominated by an i.i.d
 Bernoulli process $\tilde{X}_{i}^{k}$ with
 \begin{displaymath}
 \tilde{X}_{i}^{k} = \left\{ \begin{array}{ll}
 1 & \textrm{w.p $\alpha(n)$ }\\
 0 & \textrm{w.p $1 - \alpha(n)$}
 \end{array} \right.
 \end{displaymath} for any
 \begin{displaymath}
 \alpha(n)  \ge \frac{\textrm{maximum number of tiles touched by
 any path}}{\textrm{Total number of tiles}}.
 \end{displaymath} This follows from the observation that for any
 path $i$, $P(X_{i}^{k} = 1)$ , the probability that the $k^{th}$ tile
 is touched by the $i^{th}$ path is the same for any $k$, as the
 source/destinations are uniformly distributed in the unit square.
 From symmetry, the probability that a tile is touched by a path is
 equal to the fraction of the tiles touched by the path, which is
 $\alpha(n).$

 Noting that the number of tiles is $\frac{1}{{M(n)}^2}$, we have
 for $\alpha(n) = \frac{M(n)}{\delta}$ , $\tilde{X}_{i}^{k}$
 stochastically dominates ${X}_{i}^{k}$. We can use the above
 results to provide an upper bound on the maximum number of
 hops $H(n)$ in any tile, which  is given by
 \begin{equation}
 \label{eqn:maxhop}
 H(n) = \frac{\sqrt{2}}{\delta} \Big(\max_{k}
 \sum_{i=1}^{n} X_{i}^{k}\Big).
 \end{equation}
 \begin{claim}
 $H(n) \le \mu(n)$ \textit{almost surely}, for $\mu(n) =
 \frac{\sqrt{n\log{n}}}{\delta}
 +\sqrt{6\log{n}\frac{\sqrt{n\log{n}}}{\delta}}$.
 \end{claim}
 \begin{proof}
 \begin{eqnarray}
 \label{eqn:probound} P\Big(\max_{k} \sum_{i=1}^{n} X_{i}^{k} >
 \mu(n)\Big) &\le^{(a)} & \sum_{k=1}^{\frac{n}{\log{n}}}
 P\Big(\sum_{i=1}^{n} X_{i}^{k}> \mu(n)\Big)
 \nonumber \\
 &\le^{(b)} & \sum_{k=1}^{\frac{n}{\log{n}}}P\Big(\sum_{i=1}^{n}
 \tilde{X}_{i}^{k}> \mu(n)\Big) \nonumber \\
 &\le& n P\Big(\sum_{i=1}^{n} \tilde{X}_{i}^{k}> \mu(n)\Big)
 \end{eqnarray}
 The first inequality (a) is a union bound on the probability that
 $H(n) > \mu(n)$ and inequality (b) is due to the fact that
 $\tilde{X}$ stochastically dominates $X$. Let $\tilde{X}^{k} =
 \sum_{i=1}^{n} \tilde{X}_{i}^{k}$. Now, from \cite{motrag95}, we
 have, for sums of i.i.d bernoulli random variables that
 \begin{equation}
 \label{eqn:motbound} P\Big(\tilde{X}^{k} > (1 + \beta)
 E(\tilde{X}^{k}) \Big) \le e^{-\beta^{2}E(\tilde{X}^{k})/2}.
 \end{equation}
 We also know that $E(\tilde{X}^{k}) =
 \frac{K\sqrt{n\log{n}}}{\delta}$, from previous definitions of
 $\tilde{X}$. Taking $\beta = \sqrt{6\log{n}/E(\tilde{X}^{k})}$  in
 Equation~\ref{eqn:motbound}, we get
 \begin{equation}
 P\Big(\tilde{X}^{k} >  E(\tilde{X}^{k})
 +\sqrt{6\log{n}E(\tilde{X}^{k})} \Big) \le \frac{1}{n^{3}}.
 \end{equation}
 By summing over all $k$ in Equation~\ref{eqn:probound} we show
 that for all $\mu(n) \ge \frac{\sqrt{n\log{n}}}{\delta}
 +\sqrt{6\log{n}\frac{\sqrt{n\log{n}}}{\delta}}$, we can show that
 \begin{displaymath}
 P\Big( H(n) > \mu(n) \Big) \le \frac{1}{n^{2}}.
 \end{displaymath} The \textit{almost sure convergence} follows by
 Borel-Cantelli.
 \end{proof}

 We now outline a scheduling strategy for achieving the data-rate
 proposed. Consider a time-slot of fixed length $T$. We divide the
 time slot into $J+1$ slots of duration $\frac{T}{J+1}$ and these
 slots are allocated to tiles with the corresponding colors. Within
 the tile, the time-slots are further divided into $r(n)$
 sub-slots, where $r(n)$ is the number of hops inside the tile.
 From the previous claims, it is seen that $r(n) \le R\sqrt{n\log{n}}$
 for a large enough but finite $R$. Thus, we can guarantee that
 each hop is guaranteed a transmission time of
 $\frac{1}{R\sqrt{n\log{n}}}$ for every slot of time $T$. For any
 source-destination pair, all the hops support a data-rate of
 $\frac{1}{R\sqrt{n\log{n}}}$, and hence this throughput is achievable.
\end{proof}

\section{Conclusion}
\label{sec:conclusion} In this paper, we have presented geographic
routing strategies where the nodes have erroneous or limited
information about the destination location, and have analyzed the
asymptotic routing delays with such schemes.  Our analysis
shows that even with limited destination information (as in quadrant
routing) or erroneous angular information, the routing delays are
order-wise the same as straight-line routing.  Simulation results
indicate that the discretization effects due to node locations are
small, and there is a good match between the simulation results and
that predicted by our analysis.  We also show that for progressive
routing strategies that carry the packet closer to the destination in
each hop, the capacity is order-wise the same as a straight-line
routing strategy.

\begin{figure}
    \begin{center}
        \includegraphics[scale =.45]{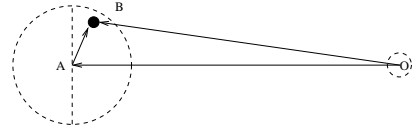}
        \caption{Geometric interpretation of the problem - Sector
          Information\label{fig:spaper}}
    \end{center}
\end{figure}

 \section*{Appendix}
 \noindent Proof of Lemma~\ref{lem:randpointbound}.
 \begin{proof}
 Consider Figure~\ref{fig:spaper}. Let B be any point inside the
 circle, and let $(S,\alpha)$ be the polar representation of the
 point. It is clear from the figure that $\overrightarrow{OB} =
 \overrightarrow{OA} + \overrightarrow{AB}.$ Now, we have
 \begin{equation}\label{eqn:basicbound1}
 |\overrightarrow{OA}| - |\overrightarrow{OB}| =
 \frac{|\overrightarrow{OA}|^2 -
 |\overrightarrow{OB}|^2}{|\overrightarrow{OA}| +
 |\overrightarrow{OB}|},
 \end{equation}where $|\overrightarrow{OA}| -
 |\overrightarrow{OB}|$ is the distance traveled towards the
 destination O in that jump. Substituting for $\overrightarrow{OB}$
 in  (\ref{eqn:basicbound1}), we obtain $|\overrightarrow{OA}| -
 |\overrightarrow{OB}|$
 \begin{equation}
 =\frac{|\overrightarrow{OA}|^2 - [ |\overrightarrow{OA}|^2 +
 |\overrightarrow{AB}|^2 -
 2|\overrightarrow{AB}||\overrightarrow{OA}|\cos{\alpha}]}{|\overrightarrow{OA}|
 + |\overrightarrow{OB}|},
 \end{equation}
 \begin{equation}
 =\frac{2S|\overrightarrow{OA}|\cos{\alpha} -
 S^2}{|\overrightarrow{OA}| + |\overrightarrow{OB}|} \label{eqn:eql-ss}
 \end{equation}
 When $|\overrightarrow{OA}|-|\overrightarrow{OB}| > 0$, we have
 \begin{equation}
 S\cos{\alpha} - \frac{S^2}{|\overrightarrow{OA}|} \le
 \frac{2S|\overrightarrow{OA}|\cos{\alpha} -
 S^2}{|\overrightarrow{OA}|+ |\overrightarrow{OB}|} =
 |\overrightarrow{OA}|-|\overrightarrow{OB}|.\label{eqn:lb1}
 \end{equation}
 The inequality in (\ref{eqn:lb1}) follows from that fact that
 $|\overrightarrow{OA}| > |\overrightarrow{OB}|,$ and by replacing
 $|\overrightarrow{OA}| + |\overrightarrow{OB}|$ with
 $2 |\overrightarrow{OA}|$ to get the lower bound.
 Next, when
 $|\overrightarrow{OA}|-|\overrightarrow{OB}| < 0$, we have
 \begin{equation}
 \frac{2S|\overrightarrow{OA}|\cos{\alpha} - S^2}{2|\overrightarrow{OA}|} \le
 \frac{2S|\overrightarrow{OA}|\cos{\alpha} -
 S^2}{|\overrightarrow{OA}|+ |\overrightarrow{OB}|}.\label{eqn:lb2}
 \end{equation}
 To see this, observe that the RHS of (\ref{eqn:lb2}) is a negative
 quantity (follows from the equality in (\ref{eqn:eql-ss})). Thus, in order to get a
 lower bound, we replace $|\overrightarrow{OA}| +
 |\overrightarrow{OB}|$ by two times the smaller of the two terms,
 i.e., $2 |\overrightarrow{OA}|$ (because in this case,
 $|\overrightarrow{OA}| \leq |\overrightarrow{OB}|$).

 Thus, as $S^2 >0$ and
 $\epsilon<|\overrightarrow{OA}|,$ from (\ref{eqn:eql-ss}),
 (\ref{eqn:lb1}) and (\ref{eqn:lb2}), we have
 \begin{equation}
 S\cos{\alpha} - \frac{S^2}{\epsilon} \le |\overrightarrow{OA}| -
 |\overrightarrow{OB}|.
 \end{equation}

 Similarly, for any point inside the circle, we can show that
 $|\overrightarrow{OA}| - |\overrightarrow{OB}| \le S\cos{\alpha}$.
 We skip the details.
 \end{proof}

\end{document}